\documentclass[%
superscriptaddress,
twocolumn,
amsmath,amssymb,
 aps,
prl,
]{revtex4-1}

\usepackage[T2A,T1]{fontenc}
\usepackage{braket}
\usepackage{graphicx}
\usepackage{graphics}
\usepackage[colorinlistoftodos]{todonotes}
\usepackage[utf8]{inputenc}
\usepackage{amsthm}
\usepackage{amsmath}
\usepackage{physics} % for bra-ket notation
\usepackage{amsfonts}
\usepackage{slashed}
\usepackage[normalem]{ulem}
\usepackage{soul}
\usepackage{graphicx,times}
\usepackage{amstext}
\usepackage{amsmath}            %serve per le subequazioni
\usepackage{amssymb}            %serve per il simbolo "marchio registrato", \circledR
\usepackage{latexsym}
\usepackage{bm}
\usepackage{color}
\usepackage{etoolbox}
\usepackage[FIGTOPCAP,raggedright,nooneline]{subfigure}
\usepackage{epstopdf}
\usepackage{lmodern}
\usepackage[english]{babel}
\usepackage{ae}
\usepackage{units}
\usepackage[americaninductors]{circuitikz}
\usepackage{tikz}
\usepackage{multirow}

\usetikzlibrary{arrows,snakes,calc}
\makeatletter
\let\cat@comma@active\@empty
\makeatother
\usepackage{url}
\usepackage[colorlinks]{hyperref}
\hypersetup{%
	plainpages=true,
	breaklinks=true,%not default in dvips mode, so we must specify
	hypertexnames=false,%not ideal, but needed when pagenums duplicate (`i' vs. `1')
	pageanchor=true,
	colorlinks=true,
	linkcolor={blue},
	citecolor={red},
	urlcolor={blue},
	anchorcolor={black}
}

\usepackage{changes}
\usepackage[footnote=true]{snotez}
\newcommand{\beq}{\begin{eqnarray}}
\newcommand{\eeq}{\end{eqnarray}}

\def\<{\langle}
\def\>{\rangle}

\def \info#1{}

\begin{document}
\title{Photon condensation and enhanced magnetism in cavity QED}% Force line breaks with \\
%\thanks{A footnote to the article title}%

\author{Juan Román-Roche}
\affiliation {Instituto de Nanociencia y Materiales de Aragón (INMA) and Departamento de Física de la Materia Condensada,
  CSIC-Universidad de Zaragoza, Zaragoza 50009,
  Spain}
  
\author{Fernando Luis}
\affiliation {Instituto de Nanociencia y Materiales de Aragón (INMA) and Departamento de Física de la Materia Condensada,
  CSIC-Universidad de Zaragoza, Zaragoza 50009,
  Spain}

\author{David Zueco}
\affiliation {Instituto de Nanociencia y Materiales de Aragón (INMA) and Departamento de Física de la Materia Condensada,
  CSIC-Universidad de Zaragoza, Zaragoza 50009,
  Spain}
%\affiliation{Fundación ARAID 50018 Zaragoza, Spain}

\date{\today}% It is always \today, today,
             %  but any date may be explicitly specified

\begin{abstract}
A system of magnetic molecules coupled to microwave cavities ($LC$ resonators) undergoes the equilibrium superradiant phase transition. The transition is experimentally observable. The effect of the coupling is first illustrated by the vacuum-induced ferromagnetic order in a quantum Ising model and then by the modification of the magnetic phase diagram of ${\rm Fe_8}$ dipolar crystals, exemplifying the cooperation between intrinsic and photon-induced spin-spin interactions.
Finally, a transmission experiment is shown to resolve the transition, measuring the quantum electrodynamical control of magnetism.
\end{abstract}

\maketitle

%%%%%%%%%%%%%%%%%%%%%%%%%%%%%%%%%%%%%%%%%%%%%%
%%%%%%%%%%%%%%%%%%%%%%%%%%%%%%%%%%%%%%%%%%%%%%
%%%%%%%%%%%%%%%%%%%%%%%%%%%%%%%%%%%%%%%%%%%%%%
%%%%%%%%%%%%%%%%%%%%%%%%%%%%%%%%%%%%%%%%%%%%%%

%%%%%%%%%%%%%%%%%%%%%%%%%%%%%%%%%%%%%%%%%%%%%%%%%%%%%
%%%%%%%%%%%%%%%%%%%%%%%%%%%%%%%%%%%%%%%%%%%%%%%%%%%%%
%%%%%%%%%%%%%%%%%%%%%%%%%%%%%%%%%%%%%%%%%%%%%%%%%%%%%
%%%%%%%%%%%%%%%%%%%%%%%%%%%%%%%%%%%%%%%%%%%%%%%%%%%%%

%{\color{red} Why is important this work? superradiance and matter modification due to q-fluc, in %particular magnetism (not explored, so far)}
%\paragraph{Introduction.-}

In 1973, Hepp and Lieb showed that $N \to \infty$ polar molecules located inside a resonant electromagnetic cavity undergo a second order transition from a normal to a superradiant phase.  The $\mathbb{Z}_2$ (parity) symmetry is spontaneously broken leading to a ferroelectric-like  state in the ``matter'' and to a nonzero population of photons in the cavity at equilibrium  \cite{Hepp1973, Hepp1973a, Wang1973}.
%Their calculation was exact in the thermodynamic ($N \to \infty$) limit \cite{Hepp1973, Hepp1973a, Wang1973}.
%
However, $47$ years and a global pandemic later, this quantum phase transition is yet to be measured \cite{Kirton2018}.
During this time, the community has enjoyed a  winding succession  of  proposals  on  how  to  achieve  the superradiant phase transition (SPT), each shortly matched by its corresponding no-go theorem \cite{Rzaewski1975, Nataf2010, counterCircuitQED, quantumhall, counterquantumhall, vuviks2014}.
Nowadays, the \emph{non-observation} of the phase transition is well understood. The crux of the matter resides in the approximations used to derive the Hamiltonian solved by Hepp and Lieb, the so-called Dicke model. 
On one side, there is the treatment of 
 the  $A^2$-term (the diamagnetic term) \cite{Garziano2020}.
On the other,  matter truncations in  different gauges must be done consistently as  pointed out by Keeling, showing that the phase transition, if any, is completely attributable to matter interactions \cite{Keeling2007}.  Consequently, the matter phase diagram remains unaltered despite it being immersed in a cavity and no photonic population develops.  The same conclusion has been recently revisited \cite{Stokes2020}.
A final step for closing the debate is found in the work by Andolina and collaborators \cite{Andolina2019}. In the thermodynamic limit $N\to \infty$, their no-go theorem illustrates how Gauge invariance inherently prohibits the SPT for electric dipoles in the long-wavelength limit. The latter implies that matter cannot respond to a static and uniform electromagnetic field.  
Therefore, to make the SPT observable, either the nature of the coupling or the spatial field distribution must be, in some way, modified.  

Theoretical proposals consider systems in the ultrastrong light-matter coupling regime \cite{DeBernardis2018} or use electron gases that either possess a Rashba spin-orbit coupling \cite{Nataf2019} or are subjected to a spatially-varying electromagnetic field \cite{Nataf2019, guerci2020, andolina2020} or have the role of photons be played by magnons \cite{bamba2020magnonic}. 
\\
Here, we propose an alternative set-up, based on magnetic molecules that couple to superconducting microwave resonators via the Zeeman interaction \cite{Jenkins2013}.
Artificial magnetic molecules \cite{gatteschi2006,bartolome2016}, designed and synthesized by chemical methods, consist of a high-spin cluster core surrounded and stabilized by a cloud of organic molecular ligands. The ability to chemically tune their relevant properties, such as the ground state spin, magnetic anisotropy and mutual interactions, combined with their stability as isolated molecular units, confer them a potential interest as magnetic memories in spintronic devices \cite{Bogani2008} and as qubits for scalable quantum information schemes \cite{Ruben2018,GaitaArino2019,Sessoli2019}. Besides, they tend to organize forming crystals, which makes them model systems to study pure magnetic dipolar order and quantum phase transitions \cite{Fernandez2000,Morello2003,Luis2005,Wen2010,Burzuri2011}.  
\\ 
This work explores the realization of the Dicke model (and generalizations of it), which undergoes the equilibrium SPT, in a crystal of molecular nanomagnets coupled to a on-chip microwave cavity. We compute the critical condition that triggers the SPT at zero and finite temperatures under rather general conditions, \emph{e.g.} for a spatially-varying magnetic field or in the presence of direct molecule-molecule interactions, and discuss a feasible method to detect it.
Besides, we build the phase diagram for the purely cavity-driven ferromagnetism and study how the spin-photon coupling enhances the intrinsic ferromagnetic order of a crystal of Fe$_{8}$ molecular clusters \cite{Burzuri2011}. 
\begin{figure}
    \centering
    \includegraphics[width = \columnwidth]{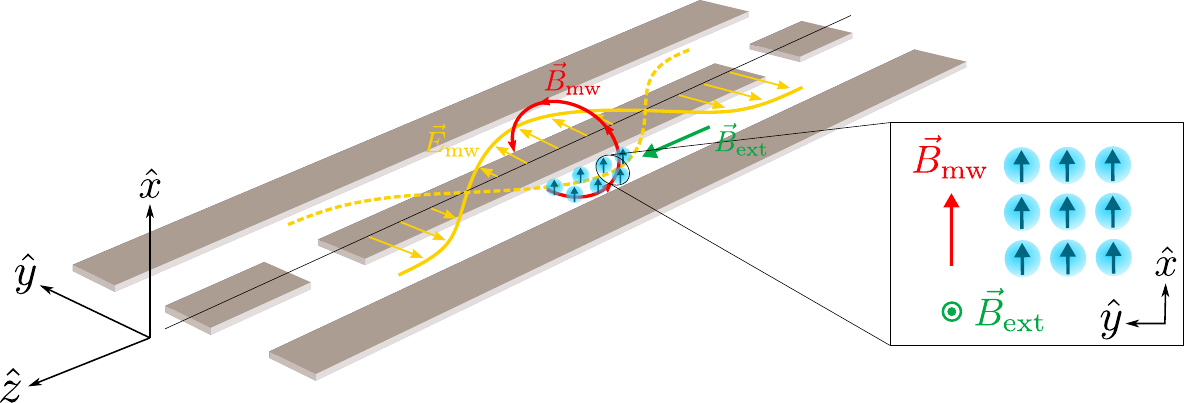}
    \caption{Schematic picture of a CPW resonator coupled to a spin ensemble. $\boldsymbol E_{\rm mw}$ and $\boldsymbol B_{\rm mw}$ are the cavity's microwave electric and magnetic fields, respectively. $\boldsymbol  B_{\rm ext}$ is the external magnetic field that induces a Zeeman splitting between the spin energy levels.}
    \label{fig:cavity fields}
\end{figure}

%%%%%%%%%%%%%%%%%%%%%%%%%%%%%%%%%%%%%%%%%%%%%%%%%%%%%
%%%%%%%%%%%%%%%%%%%%%%%%%%%%%%%%%%%%%%%%%%%%%%%%%%%%%
%%%%%%%%%%%%%%%%%%%%%%%%%%%%%%%%%%%%%%%%%%%%%%%%%%%%%
%%%%%%%%%%%%%%%%%%%%%%%%%%%%%%%%%%%%%%%%%%%%%%%%%%%%%

\paragraph{Magnetic cavity QED.-}
Hybrid platforms coupling electron \cite{Ruggenthaler2018} and, particularly, spin \cite{Xiang2013,Clerk2020} ensembles to  superconducting resonators or cavities complement circuit QED. 
Here,  the ``spins'' are superconducting qubits \cite{Wallraff2004,blais2020}. Different magnetic species have been studied in this context, including impurity spins in semiconductors \cite{Kubo2010, Schuster2010,Weichselbaumer2019}, lanthanide ions \cite{Bushev2011, Probst2014} and magnetic molecules \cite{Bonizzoni2017,Mergenthaler2017,Gimeno2020,Bonizzoni2020}. 
To observe the SPT, set ups hosting molecular crystals offer the crucial advantage of coupling a macroscopic number of identical and perfectly organized spins to a single cavity mode (cf Fig. \ref{fig:cavity fields}). 
\\
A vast majority of molecular nanomagnets are both neutral and exhibit a close to zero electric dipole. Their response to external stimuli is then accurately described by a simple 
``giant''-spin effective Hamiltonian ${\cal H}_{S}$, which includes the effects of magnetic anisotropy and the couplings to magnetic field $\boldsymbol B$ \cite{gatteschi2006,Jenkins2013}. The latter enter ${\cal H}_{S}$ via the Zeeman term $ {\mathcal H}_{\rm Z} = - g_e \mu_B \boldsymbol S \cdot \boldsymbol B$. Here $[S_i, S_j] = i \epsilon_{ijk} S_k$ are spin operators, $\mu_B = \frac{ e \hbar}{2 m}$ is the Bohr magneton and $g_e$ the Landé factor ($=2$ for an electron spin). The diamagnetic response, which arises mainly from the molecular ligands surrounding the magnetic core, is much smaller that the paramagnetic one and can be safely neglected, especially at sufficiently low temperatures.
These results provide the  experimental background for the following discussion.
\\
Let us summarize the main steps to model the molecules-cavity system in the single mode case.
For the multimode case, see \cite{SM}.
The electromagnetic field is quantized, yielding the cavity Hamiltonian $\mathcal H_{\rm c} = \hbar \Omega a^\dagger a$, with $a$ ($a^\dagger$) the photonic annihilation (creation) operators, $[a, a^\dagger]=1$. The resonance frequency $\Omega$ ranges typically between $1$ and $10$ GHz.  
The local quantized magnetic field generated by the superconducting currents can be written as
$\boldsymbol B_{\rm mw} ({\boldsymbol r}) = \boldsymbol B_{\rm rms}({\boldsymbol r}) ( a^\dagger + a) $, with
    $B_{\rm rms}^2 ({\boldsymbol r}) = \langle 0 | B_{\rm mw}^2 ({\boldsymbol r}) | 0 \rangle$
its zero-point fluctuations.

The resulting gauge-invariant cavity-spin Hamiltonian can be written as follows \cite{Jenkins2013, Jenkins2016}
\begin{align}
\nonumber
    \mathcal H &= 
     \mathcal H_S +  \mathcal H_c+  \mathcal H_I 
     \\ 
     &=
    \mathcal H_S + \hbar \Omega a^\dagger a +  \sum_j \frac{\hbar \lambda_j}{\sqrt{N}} \left( e^{i \theta_j} S^+_j  + {\rm h.c.} \right) ( a^\dagger + a) \,,
    \label{eq:HmagneticQED}
\end{align}
with $N$ the number of spins, the ladder operators $S^\pm = S^x \pm i S^y $ and the coupling constants
\begin{equation}
\label{lambdaj}
\frac{\lambda_j}{\sqrt{N}} = \frac{g_e \mu_B }{2 \hbar} |B_{\rm rms}(\boldsymbol r_j)| \, .
\end{equation}
\noindent Here, ${\boldsymbol r_j}$ is the position vector of the $j$-th spin. The phases $\theta_j$ in Eq. (\ref{eq:HmagneticQED}) are defined through  $B_{{\rm rms},x}({\boldsymbol r_j})+ i B_{{\rm rms},y}({\boldsymbol r_j}) = |B_{\rm rms}({\boldsymbol r_j})| e^{i \theta_j}$.
If the molecules are $S = 1/2$ non-interacting spins, like in a sufficiently diluted free radical sample, $\mathcal{H}_S= \hbar \frac{\omega_z}{2} \sum_j \sigma_j^z$ and the Hamiltonian \eqref{eq:HmagneticQED} matches exactly the Dicke model. 
Notice that this model does not suffer from the ``$A^2$-issue'' because the coupling is of Zeeman kind, rather than minimal (electric).  
Besides, the truncation of the electronic degrees of freedom to a finite dimensional Hilbert space is not an approximation but follows from the fact that we are dealing with ``real'' spins obeying the angular momentum commutation relations.
Both properties combined permit to avoid the no-go theorems for the SPT \cite{Andolina2019}.

%%%%%%%%%%%%%%%%%%%%%%%%%%%%%%%%%%%%%%%%%%%%%%%%%%%%%
%%%%%%%%%%%%%%%%%%%%%%%%%%%%%%%%%%%%%%%%%%%%%%%%%%%%%
%%%%%%%%%%%%%%%%%%%%%%%%%%%%%%%%%%%%%%%%%%%%%%%%%%%%%
%%%%%%%%%%%%%%%%%%%%%%%%%%%%%%%%%%%%%%%%%%%%%%%%%%%%%

\paragraph{Exact results at $N \to \infty$.-}

In order to study specific spin models and how their properties are affected by the coupling to light \emph{and vice versa}, it is convenient to obtain an effective spin Hamiltonian where the light degrees of freedom have been traced out. Following Hepp and Lieb's original derivation \cite{Hepp1973a}, this effective Hamiltonian is defined by the following expression, exact in the $N\to \infty$ limit,
\begin{equation}
    \bar Z = Tr_S\left(\frac{1}{\pi} \int d^2 \alpha \  e^{- \beta \mathcal H(\alpha)} \right) = Tr_S\left(e^{- \beta \mathcal H_{S,\text{eff}}} \right) \, ,
    \label{eq:definitionofeffectiveH} 
\end{equation}
\noindent where $\mathcal H(\alpha) = \expval{\mathcal H}{\alpha}$ and $\mathcal H$ is the total Hamiltonian given by Eq. \eqref{eq:HmagneticQED}. The resulting (See \cite{SM}) effective Hamiltonian is
\begin{equation}
    \mathcal H_{S,\text{eff}} = \mathcal H_S - \frac{1}{\hbar \Omega} \left[\sum_j \frac{\hbar \lambda_j}{\sqrt{N}}\left( e^{i \theta_j} S^+_j  + {\rm h.c.} \right) \right]^2 \, .
    \label{eq:effectiveHmag}
\end{equation}
\noindent It is apparent that the light-matter coupling translates into an effective Ising-type ferromagnetic interaction among all spins that drives the quantum phase transition. This formulation is particularly handy for studying spin models that are well captured by a mean field approach, as we show below. 

The critical point can be obtained by noticing that $[\mathcal H_I, \mathcal H_c] \sim [\mathcal H_I, \mathcal H_S] \sim 1/N$.
Then, in the thermodynamic limit, system and cavity factorize. 
Generalizing the prescription in \cite{Andolina2019} to the finite temperature case (fully developed in \cite{SM}), the critical condition can be written in  terms of the static response function $\mathcal R(T)$ of the bare spin model,
\begin{equation}
    |\mathcal R(T)| \geq \frac{\hbar \Omega}{2}   \, ,
    \label{eq:criticalcondition}
\end{equation}
where
\begin{widetext}
\begin{equation}
    \mathcal R (T) = - \frac{\sum_{m,n} e^{-\beta \epsilon_m} |\bra{\psi_m} \sum_j \frac{\hbar \lambda_j}{\sqrt{N}} \left( e^{i \theta_j} S^+_j  + {\rm h.c.} \right) \ket{\psi_n}|^2 \frac{e^{\beta \Delta_{mn}} - 1}{\Delta_{mn}}}{\sum_m e^{-\beta \epsilon_m}} \, ,
\end{equation}
\end{widetext}
with $\ket{\psi_m}$, $\epsilon_m$ the eigenstates and eigenenergies of the bare spin Hamiltonian $\mathcal H_S$ and $\Delta_{mn} = \epsilon_m - \epsilon_n$. In the case of uniform coupling, $\lambda_i \equiv \lambda$, the static response function is proportional to the magnetic susceptibility $\mathcal R = (\hbar \lambda)^2 \chi_\perp$ in the $xy$-plane, perpendicular to the external dc magnetic field (see Fig. \ref{fig:cavity fields}). A similar condition is found in three-dimensional electronic systems when the spin degrees of freedom are considered \cite{andolina2020}.

Both approaches show a relation between light and matter observables given by
\begin{equation}
    \expval{a} = \frac{1}{\hbar \Omega} \expval{\sum_j \frac{\hbar \lambda_j}{\sqrt{N}} \left( e^{i \theta_j} S^+_j  + {\rm h.c.} \right) }_S \, ,
    \label{eq:relationobservables}
\end{equation}
indicating that the phase transition is marked by the onset of both a macroscopic population of photons in the cavity and a spontaneous magnetization in the spin system.

In the case of independent spins coupled to a field pointing along $x$, we obtain the critical condition of the Dicke model for a spin $S$
\begin{equation}
    \bar \lambda^{2}=\frac{\omega_{z} \Omega}{4} \left[(2 S+1) \operatorname{cth} \left(\beta \frac{\hbar \omega_{z}}{2}(2 S+1)\right)-\operatorname{cth} \left(\beta \frac{\hbar \omega_{z}}{2}\right)\right]^{-1},
    \label{eq:critdicke}
\end{equation}
which depends on the coupling only through the collective parameter $\bar \lambda^2 \equiv N^{-1} \sum_j \lambda_j^2 $, \emph{i.e.} the root mean square of the spin-dependent couplings
\begin{equation}
  \bar \lambda^2 = V^{-1}_{\rm sample} \int_{V_{\rm sample}} \lambda^2(\boldsymbol r) dV
  =
  \frac{g_e^2 \mu_B^2 \mu_0}{8 \hbar} \rho \nu \Omega \, , 
\end{equation}
 where $V_{\rm sample}$ is the cavity volume occupied by spins and $\rho = N / V_{\rm sample}$ is the density of spins.
 In the second equality, we have used Eq. \eqref{lambdaj} and the Virial theorem and introduced the filling factor $\nu = I(V_{\rm sample}) / I(V_{\rm total})$  with $I(V) = \int_V |B_{\rm rms}(\boldsymbol r)|^2 dV$.

%%%%%%%%%%%%%%%%%%%%%%%%%%%%%%%%%%%%%%%%%%%%%%%%%%%%%
%%%%%%%%%%%%%%%%%%%%%%%%%%%%%%%%%%%%%%%%%%%%%%%%%%%%%
%%%%%%%%%%%%%%%%%%%%%%%%%%%%%%%%%%%%%%%%%%%%%%%%%%%%%
%%%%%%%%%%%%%%%%%%%%%%%%%%%%%%%%%%%%%%%%%%%%%%%%%%%%%
\begin{figure}
    \centering
    \includegraphics[width = \columnwidth]{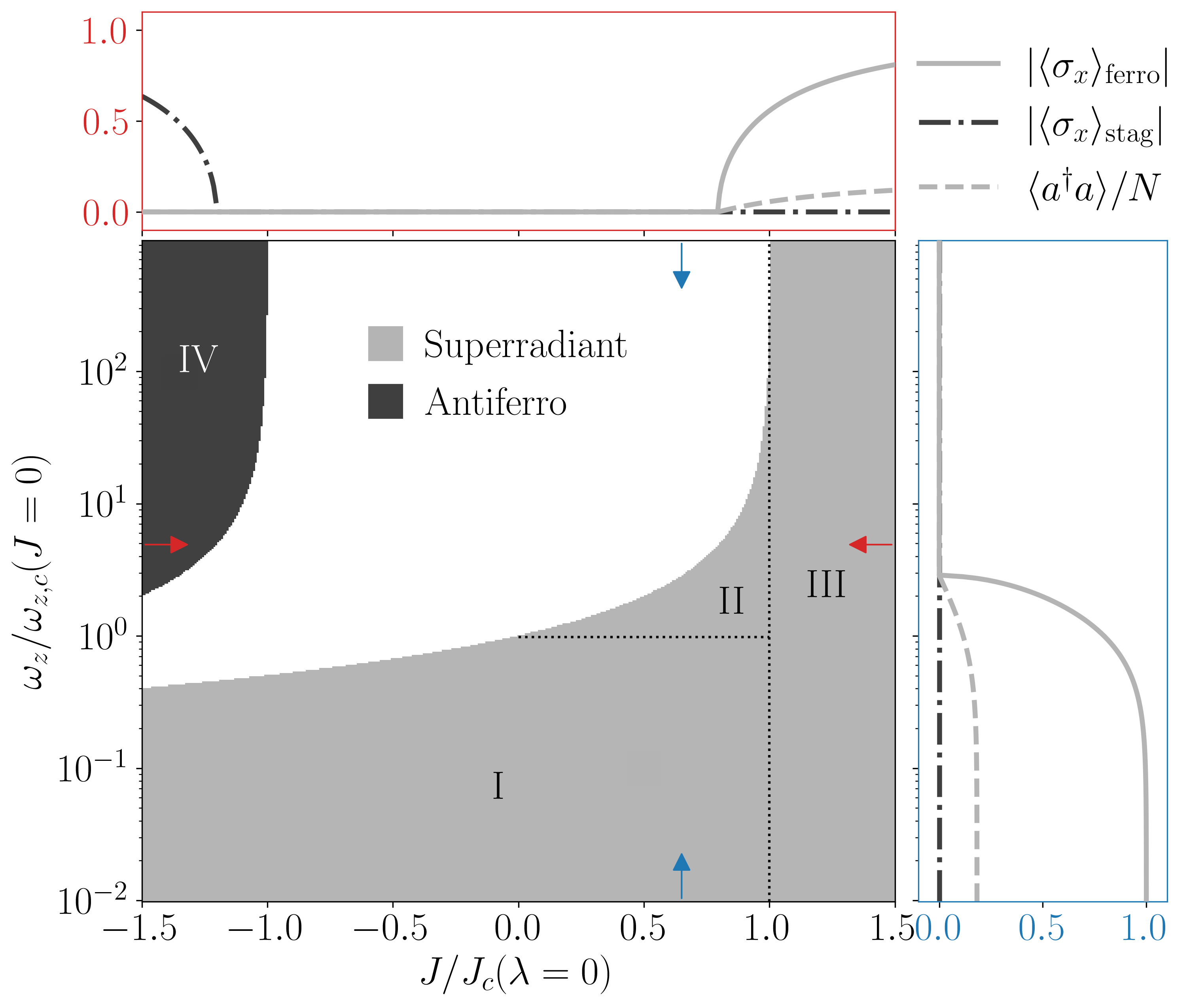}
    \caption{Mean field phase diagram of the cavity-spin Hamiltonian for $S = 1/2$ spins with ${\cal H}_S = \mathcal H_{\rm Ising}$ in 1D. The black region is antiferromagnetic. The grey region is superradiant and thus ferromagnetic. The white region is not superradiant nor magnetically ordered. For clarity, right and upper subplots showcase the standard $\expval{\sigma_x}$ and staggered $\expval{\sigma_x}_{\rm  stag}$ magnetizations, as well as the number of photons per spin $\expval{a^\dagger a}/N$, along vertical and horizontal slices of the phase diagram indicated by the arrows. For the units, $J_c (\lambda = 0)$ is the critical Ising coupling in the absence of cavity and $\omega_{z, c} (J = 0)$ is the critical spin frequency in the absence of direct Ising coupling. The parameters used are $\rho = 5.1 \cdot 10^{20} \ {\rm cm}^{-3}$, $\Omega = 1.4 \times 10^9 \ {\rm s}^{-1}$ and $\nu = 1$.}
    \label{fig:ising_phasediagram}
\end{figure}

\paragraph{Vacuum-fluctuations-driven ferromagnetism.-}
To illustrate the effects of the coupling between the cavity and the magnetic molecules as well as the interplay between intrinsic and light-induced ferromagnetic interactions, we showcase two examples using realistic experimental parameters. From a theoretical perspective, Fig. \ref{fig:ising_phasediagram} shows the alterations to the zero-temperature phases of the paradigmatic quantum Ising chain for $S = 1/2$ spins, defined by ${\cal H}_S = \mathcal H_{\rm Ising} = \frac{\hbar \omega_z}{2} \sum_j \sigma_j^z - \frac{J}{2} \sum_i \sigma_i^x \sigma_{i + 1}^x$. Note that according to Eq. \eqref{eq:relationobservables} ferromagnetic and superradiant phases are synonymous, i.e. ferromagnetic ordering is always accompanied by a finite photon population in the cavity. In particular, regions I, II and III are superradiant, as evidenced by the finite values of magnetization $|\langle \sigma_x \rangle|$ and photon number $\langle a^\dagger a \rangle / N$ shown in the insets of Fig. \ref{fig:ising_phasediagram}. Remarkably, the light-induced effective ferromagnetic interaction overpowers the intrinsic antiferromagnetic interaction extending region I into the $J < 0$ sector. In Region II it is the synergy between intrinsic and induced ferromagnetism that gives rise to superradiance. 
Finally, region III is intrinsically ferromagnetic even in the absence of the cavity and thus becomes superradiant when coupled to one. 
The use of the mean field approximation is validated in \cite{SM} by comparing the superradiant phase boundary obtained with mean field against the one obtained using exact diagonalization to compute the response function $\mathcal R$ \eqref{eq:criticalcondition}.
%{\color{red} This is the fundamental advantage of magnetic coupling: light-induced and intrinsic interactions add up, allowing us to modify the bare matter phase diagram, in stark contrast to the findings of Keeling for electric coupling \cite{Keeling2007}.} 

%%%%% experiment

\begin{figure}
    \centering
    \includegraphics[width = \columnwidth]{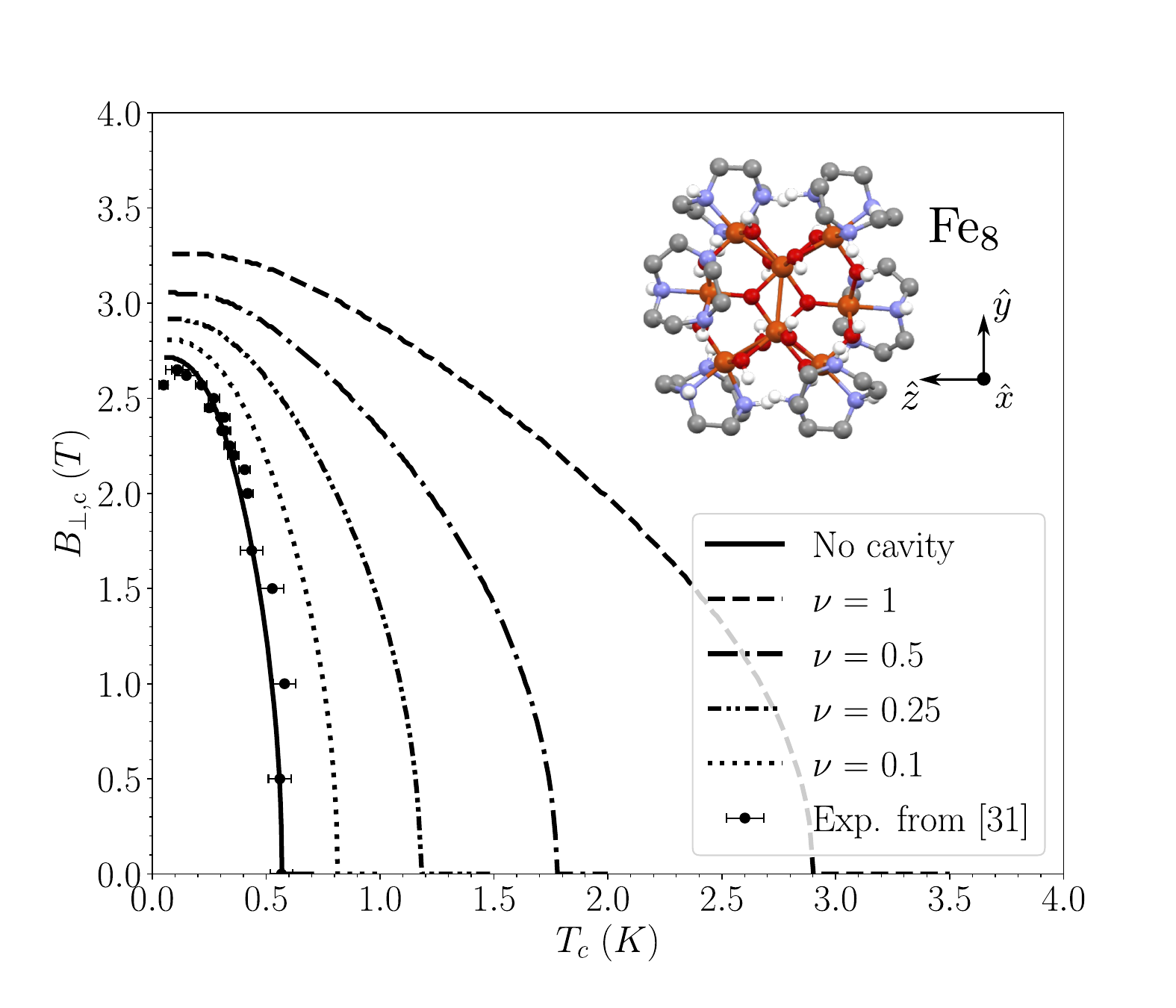}
    \caption{$B_{\rm c}-T_{\rm c}$ phase boundary determined with a quantum mean-field calculation. The solid line corresponds to the bare spin model \eqref{eq:HFe8} and is verified against experimental data (dots) obtained for a crystal of Fe$_{8}$ molecular clusters \cite{Burzuri2011}, whose structure \cite{Weighardt1984} is shown in the inset. The external field, $\boldsymbol B$, is applied perpendicular to the easy magnetization axis $x$: $\boldsymbol B = B_\perp \left (  0,  \sin \phi,-\cos \phi \right)$ with $\phi = 68^{\circ}$. The magnetic anisotropy parameters are $D/k_B = 0.294$ K and $E/k_B = 0.046$ K, and the coupling $J/k_B = 2.85 \times 10^{-3}$ K. The remaining lines are obtained with the same method taking into account the coupling to a microwave cavity for a range of filling factors $\nu$. The parameters used are $\rho = 5.1 \cdot 10^{20} \ {\rm cm}^{-3}$, taken from the crystal lattice of Fe$_{8}$ and $\Omega = 1.4 \times 10^9 \ {\rm s}^{-1}$.}
    \label{fig:modified_ising}
\end{figure}

Next, we consider a more realistic model. It corresponds to a specific molecular material, a crystal of Fe$_{8}$ clusters with $S=10$, which shows a ferromagnetic phase transition purely induced by dipolar interactions below a critical temperature $T_{\rm c} (B_\perp = 0) \cong 0.6$ K and a zero-temperature critical magnetic field $B_{\rm \perp, c} \simeq 2.65$ T \cite{Burzuri2011}. The magnetic phase diagram measured for a magnetic field perpendicular to the magnetic anisotropy axis is shown in Fig. \ref{fig:modified_ising}. As expected for a system dominated by long-range dipolar interactions, it agrees very well with the predictions of a mean-field Hamiltonian    
\begin{align}
\nonumber
    \mathcal{H}_S =& -D S_x^2 + E \left(S_z^2 - S_y^2\right) - g_e \mu_B \vec B \cdot \vec S \\
    &- 2 J \expval{S_x} S_x + J \expval{S_x}^2 \,
    \label{eq:HFe8}
\end{align}
with parameters given in Fig. \ref{fig:modified_ising}. The combination of high spin, thus high susceptibility, and negligible exchange interactions, which lead to a quite low $T_{\rm c}$ even for a densely concentrated spin lattice, makes this material well suited to obtain and measure the SPT. 

This expectation is borne out by calculations that include the coupling to a microwave cavity, whose results are also included in Fig. \ref{fig:modified_ising}. They show that light-matter interaction has a remarkable effect on the equilibrium phase diagram, enhancing both $B_{\rm \perp, c}$ and $T_{\rm c}$, the latter by almost as much as a factor six, depending on the filling of the cavity. This can be understood by noting that the effective Hamiltonian is given by Eq. \eqref{eq:HFe8} but with the coupling $J$ replaced by $J_{\rm eff} = J +  \hbar \bar \lambda^2 / \Omega$. This enhancement evidences that the cavity induces quite strong ferromagnetic correlations, a characteristic signature of the SPT, which in this case co-operate with the intrinsic interactions between the spins in the crystal. Achieving filling factors well above $0.1$ seems quite reasonable provided that one achieves a sufficiently good interface between the chip and the magnetic material.  
It can also be seen from these results that, even after the introduction of direct spin-spin interactions, the transition depends only on the filling factor, as shown in  \cite{SM}.   
We emphasize that the two previous examples show how, with magnetic coupling, the light-induced and intrinsic interactions add up, modifying the bare matter phase diagram.  This is \emph{the} signature of photon condensation and, thus, of the occurrence of the SPT   \cite{Keeling2007, Andolina2019}.
%
%\highlight{FL: is this number ok?}
%
%\highlight{FL:  Pintar las barras de error, para descartar(o no) $\nu=.1?$}

%%%%%%%%%%%%%%%%%%%%%%%%%%%%%%%%%%%%%%%%%%%%%%%%%%%%%%%%%%%%%%%%%%%%%%%%%%%%%%%%%%%%%%%%%%%%%%%%%%%%%%%%%%%%%%%%%%%%%%%%%%%%%%%%%%%%%%%%%%%%%%%%%%

\paragraph{Transmission experiment for resolving the transition.-} 

Finally, we discuss how to measure a signature of the phase transition.
A direct route would be to measure the order temperature of the magnetic material \emph{inside} and \emph{outside} the cavity.  
However, conventional methods to measure $T_{\rm c}$ (or $B_{\rm c}$), based on magnetic susceptibility or neutron diffraction \cite{Burzuri2011}, do not lend themselves easily to include a superconducting cavity. Besides, they require very large crystals, often much larger than the typical cavity volumes.
Therefore, we envision here a more accessible way: a  transmission experiment, where the cavity is coupled to a microwave transmission line. A signal, sent through it, interacts with the cavity-spins system and the transmitted signal is recorded.
Since this signal is proportional to the dynamical response of the system, we expect to observe a signature  near the transition.
Technically, the calculation involves computing the dynamical susceptibility of the whole system (cavity plus spins) using a quantum master equation.
Typically, the dissipation for both the spins and the cavity is added phenomenologically, inserting the terms as if the cavity and the spins were not coupled. However, this, so-called, \emph{local} approach has been criticized. A rigorous derivation involves taking into account the coupling between the two subsystems, obtaining the global master equation. The differences between both approaches are relevant in strongly coupled systems \cite{Hofer_2017, Carusotto2006, Gonzlez2017}. Since we are interested in what happens near the transition, we explore the differences between local and global approaches to rule out that the signature is an artifact of the approximations taken.
We notice, however, that in the global approach both eigenvectors and eigenvalues for the coupled system are needed.
These are impossible to obtain in a full treatment.
On the other hand, if we consider a spin $1/2$ Dicke model with homogeneous coupling to the cavity a Holstein-Primakoff transformation allows us to write the total Hamiltonian as two coupled oscillators.   
Then, the system is exactly solvable and the global master equation can be obtained.
By doing so, we can compare both local and global approaches for resolving the transition.

\begin{figure}
    \centering
    \includegraphics[width = \columnwidth]{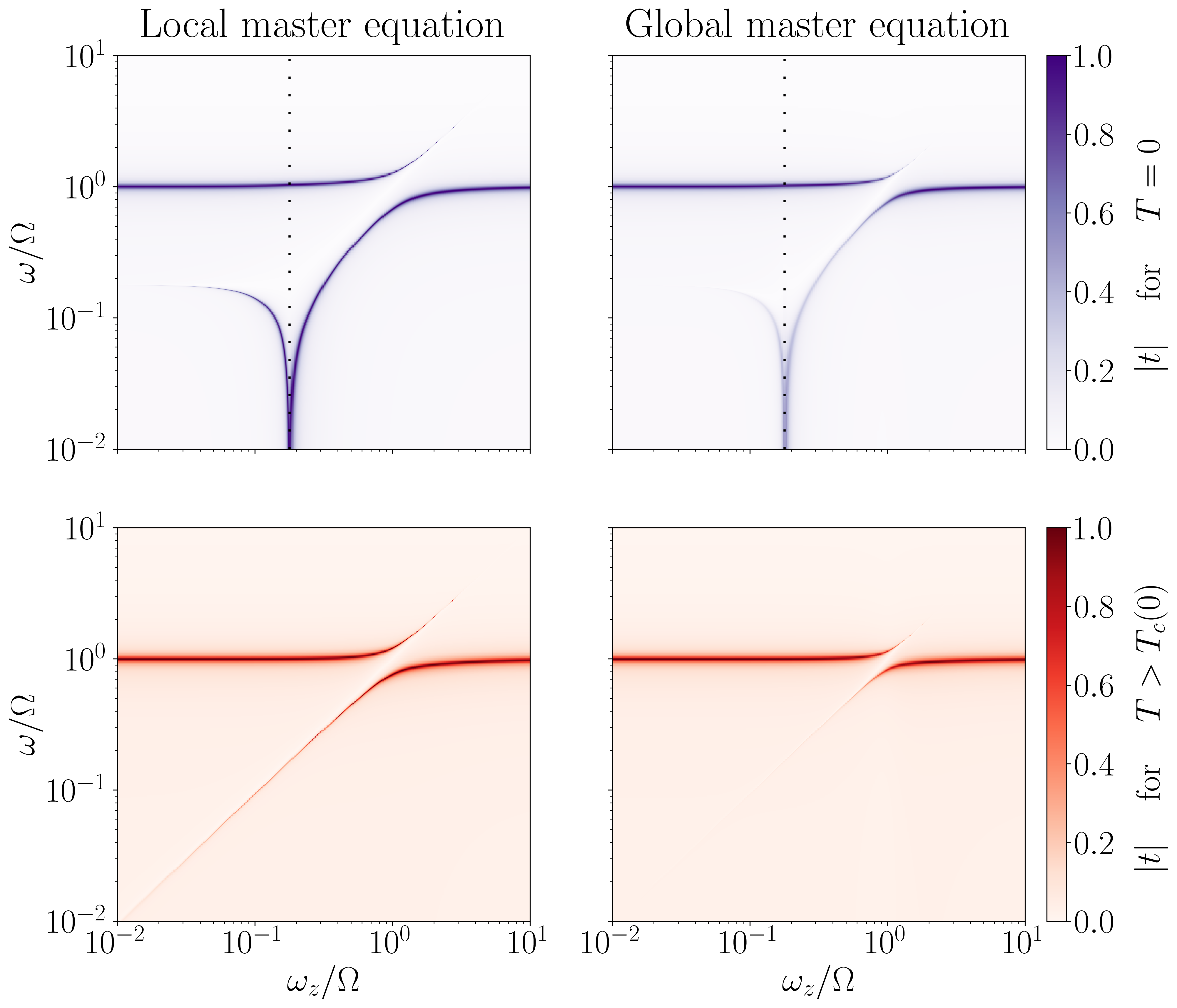}
    \caption{ Plots of the transmission for temperatures below and above the critical temperature at zero field (top and bottom) computed with the local and global master equation (left and right) as a function of the ratios between $\omega$, $\omega_z$ and $\Omega$ for a spin 1/2 Dicke model. A dotted black line marks the transition to the superradiant phase at $\omega_{\rm z, c} (0)$. The parameters used are  $\gamma_1 = 0.025 \, \Omega$, $\gamma_2 = 0.025 \, \omega_z$, $ \rho = 5.1 \times 10^{20} \, {\rm cm}^{-3}$, $\Omega = 1.4 \times 10^9 \ {\rm s}^{-1}$ and $\nu = 0.25$.}
    \label{fig:trasnmission}
\end{figure}

The explicit formulas are rather involved and thus sent to \cite{SM}.
In figure  \ref{fig:trasnmission} we summarize our findings.
We show 2D transmission plots for temperatures larger or smaller than the zero-field critical temperature $T_{\rm c} (0)$. In the former limit, the system remains disordered, regardless of $\omega_z$ and only shows the well-known avoided crossing at $\omega = \Omega = \omega_z$. As temperature is decreased below $T_{\rm c}(0)$ the magnetic and photon states depend on $\omega_z$. The phase transition, for $\omega_z \sim \omega_{\rm z, c} (0)$, is then characterized by the appearance of a new resonance, i.e. a new transmission channel, at $\omega_z \sim \omega_{\rm z, c} (0)$.
On physical grounds, this is a consequence of the vanishing frequency of the lower mode at the transition, which increases the response at equilibrium ($\omega \to 0$).
Besides, the resonance appears in both the local and global approaches, supporting our findings.
Quantitative differences appear, obviously, but this is expected
since one of the pitfalls of the local approach is not reproducing the correct equilibrium states.
This behaviour constitutes a clear signature of the SPT and shows that this transition can be detected in a standard temperature-dependent transmission experiment, provided that the spin-photon coupling is large enough.
%%

%%%%%%%%%%%%%%%%%%%%%%%%%
%%%%%%%%%%%%%%%%%%%%%%%%%
%%%%%%%%%%%%%%%%%%%%%%%%%
%%%%%%%%%%%%%%%%%%%%%%%%%
\paragraph{Conclusions.-}

We have shown that the coupling of a macroscopic number of spins, in particular crystals of molecular nanomagnets, to the quantum vacuum fluctuations of cavities or $LC$ resonators generates ferromagnetic spin-spin interactions that lead to an equilibrium superradiant phase transition. Using realistic parameters, we find that it also gives rise to detectable signatures in the transmission of microwaves through such a hybrid set-up, thus providing a feasible solution to the long-standing problem of measuring the SPT.
Our results also present these systems as ideal for exploring the quantum electrodynamical control of matter, baptized as \emph{cavity QED materials} \cite{rokaj2020free}. Recent studies have shown that quantum light fluctuations can modify properties such as excitonic transport \cite{Feist2015, Schachenmayer2015,Orgiu2015}, chemical reactivity  \cite{Thomas2016, Thomas2019}, superconductivity \cite{Sentef2018, Schlawin2019, Curtis2019, thomas2019b} and the ferroelectric phase in quantum paramagnetic materials \cite{Pilar2020, Schuler2020, ashida2020}. Here, we show that they can also generate, modify and control long-range ordered magnetic phases.

%%%%%%%%%%%%%%%%%%%%%%%%%
%%%%%%%%%%%%%%%%%%%%%%%%%
%%%%%%%%%%%%%%%%%%%%%%%%%
%%%%%%%%%%%%%%%%%%%%%%%%%
\paragraph{Acknowledgments.-}
The authors acknowledge funding from the EU (COST Action 15128 MOLSPIN, QUANTERA SUMO and FET-OPEN Grant 862893 FATMOLS), the Spanish MICINN (MAT2017-88358-C3-1-R, RTI2018-096075-B-C21, PCI2018-093116, EUR2019-103823), the Gobierno de Arag\'on (Grant E09-17R Q-MAD) and the BBVA foundation (Leonardo Grants 2018).
%========== Bibliography =============
\bibliography{main.bib}

\end{document}

\title{Supplemental Material for \\ ``Photon Condensation and Enhanced Magnetism in Cavity QED''}% Force line breaks with \\
%\thanks{A footnote to the article title}%

\author{Juan Román-Roche}
\affiliation {Instituto de Nanociencia y Materiales de Aragón (INMA) and Departamento de Física de la Materia Condensada,
  CSIC-Universidad de Zaragoza, Zaragoza 50009,
  Spain}
  
 \author{Fernando Luis}
\affiliation {Instituto de Nanociencia y Materiales de Aragón (INMA) and Departamento de Física de la Materia Condensada,
  CSIC-Universidad de Zaragoza, Zaragoza 50009,
  Spain}
  
\author{David Zueco}
\affiliation {Instituto de Nanociencia y Materiales de Aragón (INMA) and Departamento de Física de la Materia Condensada,
  CSIC-Universidad de Zaragoza, Zaragoza 50009,
  Spain}

\date{\today}% It is always \today, today,
             %  but any date may be explicitly specified

\begin{abstract}
In this Supplemental Material we provide additional information on the modelling of the interaction between light and molecules (Sec. \ref{light-mol_interact}), the effective spin Hamiltonian (Sec. \ref{effectivehamiltonian}), the critical condition in terms of the spin static response function (Sec. \ref{criticalcondition}), the finite temperature stiffness theorem (Sec. \ref{stiffnessfiniteT}), the calculation of the modified Ising phase diagram (Sec. \ref{comparisonphasediagram}), the dependence of the effective spin-spin interaction on $\bar \lambda$ (Sec. \ref{rootmeansquare}) and the transmission experiment (Sec. \ref{transmission}).
\end{abstract}

\maketitle
\section{Modeling the interaction of light and magnetic molecules}
\label{light-mol_interact}

If we consider the source of the electromagnetic field to be an optical or superconducting cavity, we quantize the electromagnetic field in the Coulomb gauge, where $\boldsymbol A(\boldsymbol r_i) = \sum_k \boldsymbol A_k(\boldsymbol r_i) (a_k + a_k^\dagger )$, $\boldsymbol B(\boldsymbol r_i) = \boldsymbol \nabla \times \boldsymbol A(\boldsymbol r_i)$ and $a_k, a_k^\dagger$ are the bosonic annihilation and creation operators. The interaction of the spins with the magnetic field through Zeeman coupling $ {\mathcal H}_{\rm Z} = - g_e \mu_B  \boldsymbol S \cdot \boldsymbol B$ can be split into two contributions, one from the external field inducing level splitting on the spin energy levels which is included in $\mathcal H_S$, and another from the microwave field generated by the cavity which constitutes $\mathcal H_I$. The local multimode quantized magnetic field generated by the superconducting currents can be written as $\boldsymbol B_{\rm mw} ({\boldsymbol r}) = \sum_k \boldsymbol B_{k, \rm rms}({\boldsymbol r}) ( a_k^\dagger + a_k) $, where we have used the orthogonality of the electromagnetic modes to write $B_{\rm rms}^2 ({\boldsymbol r}) =  \langle 0 | B_{\rm mw}^2 ({\boldsymbol r}) | 0 \rangle = \sum_k B_{k, \rm rms}^2({\boldsymbol r})$. The  gauge  invariant  full  cavity-spin Hamiltonian is in general
\begin{equation}
    \mathcal H = \mathcal H_S + \sum_k \hbar \Omega_k a_k^\dagger a_k +  \sum_k \sum_j \frac{\hbar \lambda_{j, k}}{\sqrt{N}} \left( e^{i \theta_j} S^+_j  + {\rm h.c.} \right) ( a_k^\dagger + a_k) \,,
\end{equation}
where
\begin{equation}
\frac{\lambda_{j,k}}{\sqrt{N}} = \frac{g_e \mu_B }{2 \hbar} |B_{k,\rm rms}(\boldsymbol r_j)| \, .
\end{equation}
\section{Computing the effective spin Hamiltonian}
\label{effectivehamiltonian}

Despite the fact that the bounds for the partition function were essentially proven by Hepp and Lieb \cite{Hepp1973, Lieb1973}, let us prove it again here for self-consistency. We want to show that
\begin{equation}
    \bar Z \leq Z \leq e^{\beta \hbar \Omega} \bar Z; \qquad \text{with} \qquad \bar Z = Tr_S \left(\frac{1}{\pi} \int d^2 \alpha \  e^{-\beta \mathcal H(\alpha)}\right) \, .
    \label{eq:Zbounds}
\end{equation}
where $\mathcal H(\alpha) = \expval{\mathcal H}{\alpha}$. Consider a Hamiltonian of the form 
\begin{equation}
    \mathcal H =
    \mathcal H_S + \hbar \Omega a^\dagger a +  \sum_j \frac{\hbar \lambda_j}{\sqrt{N}} \left( e^{i \theta_j} S^+_j  + {\rm h.c.} \right) ( a^\dagger + a) \; .
\end{equation}
In essence, the only relevant details are that the light-matter coupling is linear in the bosonic operators and that the only non linear contribution is trough the photonic term $\hbar \Omega a^\dagger a$. We will start with the lower bound. For that, we make use of Jensen's inequality for convex functions, which states that, for a convex function $f$
\begin{equation}
    f\left(\sum_i \lambda_i x_i\right) \leq \sum \lambda_i f(x_i) \, .
\end{equation}
Because the exponential is a convex function, it follows that
\begin{align}
     Z =& Tr_S \left(\frac{1}{\pi} \int d^2 \alpha \  \bra{\alpha} e^{-\beta \mathcal H}  \ket{\alpha} \right) \geq \notag \\ &Tr_S \left(\frac{1}{\pi} \int d^2 \alpha \   e^{-\beta \bra{\alpha}\mathcal H\ket{\alpha}}   \right)  = Tr_S \left(\frac{1}{\pi} \int d^2 \alpha \  e^{-\beta \mathcal H(\alpha)}\right) =  \bar Z \, .
\end{align}
The upper bound requires some more algebra. We will make use of the Golden-Thompson inequality: let $A$ and $B$ be Hermitian operators, then
\begin{equation}
    Tr\left((AB)^n\right) \leq Tr\left(A^n B^n\right) \, .
\end{equation}
We are concerned with obtaining an upper bound for $Z$, which we can express as
\begin{equation}
    Z = Tr \left(e^{-\beta \mathcal H} \right) = Tr \left(\left(1 - \frac{\beta}{m} \mathcal H \right)^m \right) \, ,
\end{equation}
for $m \to \infty$. Before advancing further, it is convenient to express $a^\dagger a = a a^\dagger - 1$ in terms of a sum of projectors of coherent states, using the closure relation $\pi^{-1} \int d^2 \alpha \ket{\alpha}\bra{\alpha}$,
\begin{equation}
    a^\dagger a = \frac{1}{\pi} \int d^2 \alpha \left( a \ket{\alpha}\bra{\alpha} a^\dagger - \ket{\alpha}\bra{\alpha} \right) = \frac{1}{\pi} \int d^2 \alpha  \left( |\alpha|^2 -  1\right)\ket{\alpha}\bra{\alpha} \, .
\end{equation}
We can now write
\begin{equation}
    1 - \frac{\beta}{m} \mathcal H =  1 - \frac{1}{\pi} \int d^2 \alpha  \left( \mathcal H(\alpha) -  \hbar \Omega\right)\ket{\alpha}\bra{\alpha} = \frac{1}{\pi} \int d^2 \alpha \left(1 - \frac{\beta}{m}  \left( \mathcal H(\alpha) -  \hbar\Omega\right)\right) \ket{\alpha}\bra{\alpha} \, .
\end{equation}
Defining the multiplication operator $A$
\begin{equation}
    A \ket{\alpha} = \left(1 - \frac{\beta}{m}  \left( \mathcal H(\alpha) -  \hbar \Omega\right)\right) \ket{\alpha} \, ,
\end{equation}
we can write the previous expression as
\begin{equation}
    1 - \frac{\beta}{m} \mathcal H = A \frac{1}{\pi} \int d^2 \alpha \ket{\alpha}\bra{\alpha} = A B \, .
\end{equation}
At this point we use the Golden-Thompson inequality, noticing that $B^n = \mathbb{I}$
\begin{equation}
    Tr\left(\left(1 - \frac{\beta}{m} \mathcal H \right)^m \right) = Tr\left((AB)^m\right) \leq Tr\left(A^m B^m\right) = Tr\left(A^m\right) \, ,
\end{equation}%
which finally gives us the upper bound
\begin{equation}
    Z \leq Tr\left(\left(1 - \frac{\beta}{m}  \left( \mathcal H(\alpha) -  \hbar \Omega \right)\right)^m\right) = Tr\left(e^{- \beta \left( \mathcal H(\alpha) -  \hbar \Omega\right)} \right) = e^{\beta \hbar \Omega} \bar Z \, .
\end{equation}
It can be readily seen that the generalization to a multimode field simply yields $Z \leq \exp[\beta \sum_k \hbar \Omega_k] \bar Z$. Provided the number of modes is finite or alternatively that $\sum_k \hbar \Omega_k$ is, the bounds for $Z$ become exact equalities in the thermodynamic limit. 

At this point, we define the effective spin Hamiltonian through the following relation
\begin{equation}
    \bar Z = Tr_S\left(\frac{1}{\pi} \int d^2 \alpha \  e^{- \beta \mathcal H(\alpha)} \right) = Tr_S\left(e^{- \beta \mathcal H_{S,\text{eff}}} \right) \, ,
    \label{eq:definitionofeffectiveH} 
\end{equation}
Where, for a general spin Hamiltonian, we define $\mathcal H(\alpha)$ as
\begin{equation}
    \mathcal H(\alpha)  = \mathcal H_S + \hbar \Omega |\alpha |^2 + \sum_j \frac{\hbar \lambda_j}{\sqrt{N}} \left( e^{i \theta_j} S^+_j  + {\rm h.c.} \right) (\alpha^* + \alpha) \, .
\end{equation}
Thus
\begin{equation}
    \frac{1}{\pi} \int d^2 \alpha \  e^{- \beta \mathcal H(\alpha)} \propto \frac{1}{\pi} \int d\Re(\alpha) \  e^{-\beta\left(\mathcal H_S + \hbar \Omega \Re(\alpha)^2 + \sum_j \frac{\hbar \lambda_j}{\sqrt{N}} \left( e^{i \theta_j} S^+_j  + {\rm h.c.} \right) \Re(\alpha)  \right)} \, .
\end{equation}
Which yields
\begin{equation}
    \mathcal H_{S,\text{eff}} = \mathcal H_S - \frac{1}{\hbar \Omega} \left[\sum_j \frac{\hbar \lambda_j}{\sqrt{N}}\left( e^{i \theta_j} S^+_j  + {\rm h.c.} \right) \right]^2 \, .
    \label{eq:effectiveHmag}
\end{equation}
We have used the fact that $[\mathcal H_S / N, \sum_j \frac{\hbar \lambda_j}{\sqrt{N}} \left( e^{i \theta_j} S^+_j  + {\rm h.c.} \right) / N] \to 0$ in the thermodynamic limit, provided that $\mathcal H_S$ is extensive, in order to extract the exponential of $\mathcal H_S$ for integration and then to put together $\mathcal H_{\text{S,eff}}$.

\section{Computing the critical condition in terms of the spin static response function}
\label{criticalcondition}
Let us decompose the Hamiltonian as $\mathcal H = \mathcal H_{\rm ph} + \mathcal H_{S} + \mathcal H_{\rm int}$. Where $\mathcal H_S$ is model dependent and 
\begin{align}
    &\mathcal H_{\rm ph} = \hbar \Omega a^\dagger a \, , \\
    &\mathcal H_{\rm int} = \sum_j \frac{\hbar \lambda_j}{\sqrt{N}} \left( e^{i \theta_j} S^+_j  + {\rm h.c.} \right) (a^\dagger + a) \, .
\end{align}
We found that a generalization of Andolina and colleagues' approach \cite{Andolina2019} is convenient for dealing with general Hamiltonians of this sort. Let us first consider the zero temperature case, and then advance to finite temperature.

\subsection{Zero temperature $T = 0$}

In the thermodynamic limit, $N \to \infty$, it can be shown that $[\mathcal H_{ph} / N, \mathcal H_{int} / N] = [\mathcal H_S / N, \mathcal H_{int} / N] = 0$, thus light and spin degrees of freedom are effectively decoupled and the states of the system are separable. In such scenario a mean-field treatment becomes exact, and we can propose for the photons in the cavity a coherent state $a \ket{\alpha} = \alpha \ket{\alpha}$. Taking $\alpha \in \mathbb{R}$ and a spin state $\ket{\psi}$, we find
\begin{equation}
    E_\psi(\alpha) = \hbar \Omega \alpha^2 + \expval{\mathcal H_S}{\psi} + 2 \alpha \expval{\sum_j \frac{\hbar \lambda_j}{\sqrt{N}} \left( e^{i \theta_j} S^+_j  + {\rm h.c.} \right)}{\psi} \, .
    \label{eq:Epsi}
\end{equation}
Minimizing with respect to $\alpha$ yields
\begin{equation}
    \expval{\sum_j \frac{\hbar \lambda_j}{\sqrt{N}} \left( e^{i \theta_j} S^+_j  + {\rm h.c.} \right)}{\psi} = - \hbar \Omega \alpha  \, ,
    \label{eq:constrain}
\end{equation}
which gives a constraint in the search for the ground state of $\mathcal H_S$. Introducing Eq. \eqref{eq:constrain} in Eq. \eqref{eq:Epsi} gives the minimum energy for a state with non-zero $\alpha$
\begin{equation}
    E_\psi(\alpha) = \expval{\mathcal H_S}{\psi} - \hbar \Omega \alpha^2 \, .
\end{equation}
For the state $\ket{\Psi} = \ket{\psi} \otimes \ket{\alpha}$ with $\alpha \neq 0$ to be the true minimum requires $\ket{\psi}$ to be the least energetic state obeying the constraint \eqref{eq:constrain} (termed \textit{constrained ground state}) and, of course, $E_\psi(\alpha) \leq E_{\psi_0}(0)$, i.e.
\begin{equation}
    \expval{\mathcal H_S}{\psi} - \expval{\mathcal H_S}{\psi_0} \leq \hbar \Omega \alpha^2 \, .
    \label{eq:sptcondition}
\end{equation}
Where $\ket{\psi_0}$ is the unconstrained ground state of $\mathcal H_S$. The stiffness theorem \cite{Tosatti2006} allows us to expand $\expval{\mathcal H_S}{\psi} - \expval{\mathcal H_S}{\psi_0}$ in powers of $\alpha$. Assuming that $\expval{\sum_j \frac{\hbar \lambda_j}{\sqrt{N}} \left( e^{i \theta_j} S^+_j  + {\rm h.c.} \right)}{\psi_0} = 0$, we find
\begin{equation}
    \expval{\mathcal H_S}{\psi} - \expval{\mathcal H_S}{\psi_0} = - \frac{\alpha^2}{2 \tilde{\mathcal R}} \, ,
    \label{eq:stiffness}
\end{equation}
with $\tilde{\mathcal R}$ being the static response function
\begin{equation}
    \tilde{\mathcal R} = - \frac{2}{(\hbar \Omega)^2} \sum_{m \neq 0} \frac{|\bra{\psi_m} \sum_j \frac{\hbar \lambda_j}{\sqrt{N}} \left( e^{i \theta_j} S^+_j  + {\rm h.c.} \right) \ket{\psi_0}|^2}{\epsilon_m - \epsilon_0} \, .
    \label{eq:response}
\end{equation}
$\ket{\psi_m}$ and $\epsilon_m$ are respectively the eigenstates and eigenenergies of $\mathcal H_S$ such that $\mathcal H_S \ket{\psi_m} = \epsilon_m \ket{\psi_m}$. From Eqs. \eqref{eq:sptcondition}-\eqref{eq:response} we obtain the condition for the superradiant phase transition of a general spin Hamiltonian $\mathcal H_S$
\begin{equation}
    \frac{\hbar \Omega}{4} \leq \sum_{m \neq 0} \frac{|\bra{\psi_m} \sum_j \frac{\hbar \lambda_j}{\sqrt{N}} \left( e^{i \theta_j} S^+_j  + {\rm h.c.} \right) \ket{\psi_0}|^2}{\epsilon_m - \epsilon_0} \, .
    \label{eq:responsecondition}
\end{equation}
Note that the static response function used in the main text $\mathcal R$ is defined as $\mathcal R = (\hbar \Omega)^2 \tilde{\mathcal R}$, where $\tilde{\mathcal R}$ is the response function defined here above.
\subsection{Finite temperature $T \geq 0$}
Firstly, let us show that minimizing the energy is either sufficiently or completely equivalent to minimizing the free energy at low temperatures.
The free energy is defined as $F = E - T S$. Consider now a variation in $F$ at constant temperature
\begin{equation}
    \Delta F = \Delta E - T \Delta S \, .
\end{equation}
Clearly, at $T = 0$, $\Delta  F = \Delta E$, so minimizing the free energy is equivalent to minimizing the energy at $T=0$. For higher temperatures, we can write
\begin{equation}
    \frac{\Delta E - \Delta F}{T} = \Delta S \, .
\end{equation}
Taking the low temperature limit we have
\begin{equation}
    \lim_{T \to 0} = \frac{\Delta E - \Delta F}{T} = \lim_{T \to 0} \Delta S \, .
\end{equation}
Applying L'Hopital's rule and Nernst postulate $\lim_{T \to 0} \Delta S  = 0$, we find
\begin{equation}
    \left(\frac{d \Delta E}{d T} \right)_{T= 0} - \left(\frac{d \Delta F}{d T} \right)_{T= 0} = 0 \, .
\end{equation}
Therefore, not only are $\Delta E$ and $\Delta F$ equal at $T = 0$ but so are their slopes, ensuring that $\Delta E$ and $\Delta F$ remain equal in a range of low temperatures, and not only at strictly $T = 0$. This implies that the minimization of $F$ and of $E$ become equivalent at low temperatures.
Now, if we consider that our thermal spin ensemble state is captured by the density matrix $\rho$, the energy of the system is
\begin{equation}
    E_\rho(\alpha) = \hbar \Omega \alpha^2 + Tr\left(\mathcal H_S \rho \right) + 2 \alpha \ Tr\left(\sum_j \frac{\hbar \lambda_j}{\sqrt{N}} \left( e^{i \theta_j} S^+_j  + {\rm h.c.} \right) \rho\right) \, ,
\end{equation}
which we can minimize with respect to $\alpha$ to yield a constrained minimum condition for the spin system
\begin{equation}
    Tr\left(\sum_j \frac{\hbar \lambda_j}{\sqrt{N}} \left( e^{i \theta_j} S^+_j  + {\rm h.c.} \right) \rho \right) = \expval{\sum_j \frac{\hbar \lambda_j}{\sqrt{N}} \left( e^{i \theta_j} S^+_j  + {\rm h.c.} \right)} = - \hbar \Omega \alpha \, .
\end{equation}
Reintroducing it in our expression for the energy gives
\begin{equation}
    E_\rho(\alpha) = Tr(\mathcal H_S \rho) - \hbar \Omega \alpha^2 \, .
\end{equation}
With that, the critical condition becomes $E_\rho(\alpha) \leq E_{\rho_0}(0)$, which amounts to %
\begin{equation}
    Tr(\mathcal H_S \rho) - Tr(\mathcal H_S \rho_0) \leq \hbar \Omega \alpha^2 \, .
    \label{eq:minimumconfiniteT}
\end{equation}
We can now make use of the extended stiffness theorem (See Sec. \ref{stiffnessfiniteT}) to write the left hand side of Eq. \eqref{eq:minimumconfiniteT} in terms of powers of $\alpha$ and then obtain a critical condition analogous to the zero temperature case, but with a static response function that depends on temperature
\begin{equation}
    -\frac{1}{2} \frac{\alpha^2}{\tilde{\mathcal R}(T)} \leq \hbar \Omega \alpha^2 \to -\tilde{\mathcal R}(T) \geq \frac{1}{2 \hbar \Omega} \, .
\end{equation}
Considering the temperature dependent static response function we obtain the critical condition
\begin{equation}
   \frac{\hbar \Omega}{2} \leq \frac{\sum_{m,n} e^{-\beta \epsilon_m} |\bra{\psi_m} \sum_j \frac{\hbar \lambda_j}{\sqrt{N}} \left( e^{i \theta_j} S^+_j  + {\rm h.c.} \right) \ket{\psi_n}|^2 \frac{e^{\beta \Delta_{mn}} - 1}{\Delta_{mn}}}{\sum_m e^{-\beta \epsilon_m}} \, .
   \label{eq:criticalcondofT}
\end{equation}
Where $\ket{\psi_m}$ and $\epsilon_m$ are respectively the eigenstates and eigenenergies of $\mathcal H_S$ and $\Delta_{mn} = \epsilon_m - \epsilon_n$.
\section{Extending the Stiffness theorem to finite temperatures.}
\label{stiffnessfiniteT}

Let us extend the Stiffness theorem \cite{Tosatti2006} to deal with finite temperatures. We consider a constrained minimization problem, with the energy given by
\begin{equation}
    E(A) = Tr(\mathcal{\hat H} \rho_A) \, ,
\end{equation}
where $\rho_A$ indicates that $A = \langle \hat A \rangle = Tr(\hat A \rho_A)$. We now construct the Hamiltonian $\mathcal{\hat H}_A = \mathcal{\hat H} + F_A \hat A$ such that
\begin{equation}
    \rho_A = \frac{1}{Z_A} e^{-\beta(\mathcal{\hat H} + F_A \hat A)} \, .
\end{equation}
By the definition of the static response function, we impose
\begin{equation}
    A = \langle \hat A \rangle = \langle \hat A \rangle_0 + F_A \mathcal R_{AA(T)} \implies F_A = \frac{A}{\mathcal R_{AA}(T)} \, .
\end{equation}
With that, we can express
\begin{equation}
    E(A) = Tr(\mathcal{\hat H}_A \rho_A) - F_A A \, ,
    \label{eq:restrictedenergyT}
\end{equation}
which we now wish to compute. For that purpose, consider a parametrized Hamiltonian with parameter $0 \leq \lambda \leq 1$,
\begin{equation}
    \mathcal{\hat H}_A(\lambda) = \mathcal{\hat H} + \lambda F_A \hat A \, .
\end{equation}
Extending the equations presented thus far, we find
\begin{equation}
    A(\lambda) = Tr(\hat A \rho_A(\lambda)) = \lambda F_A \mathcal R_{AA}(T) \, ,
\end{equation}
with
\begin{equation}
    \rho_A(\lambda) = \frac{1}{Z_A(\lambda)} e^{-\beta \mathcal{\hat H}_A(\lambda)} \, .
\end{equation}
Applying the Hellman-Feynman theorem again we find
\begin{equation}
    \begin{split}
        E_A(1) &= E(0) + \int_0^1 d\lambda  Tr(\frac{\partial \mathcal{\hat H}_A(\lambda)}{\partial \lambda}) = \\
        &= E(0) + \int_0^1 d\lambda \lambda \mathcal R_{AA}(T) F_A^2 = \\
        &= E(0) + \frac{1}{2} \mathcal R_{AA}(T) F_A^2 \, .
    \end{split}
\end{equation}
Substituting back into Eq. \eqref{eq:restrictedenergyT} we finally obtain
\begin{equation}
    E(A) = E(0) - \frac{1}{2}\frac{A^2}{\mathcal R_{AA}(T)} \, .
\end{equation}
So the result is analogous to the zero temperature case with the only difference that the static response function $\mathcal R_{AA}(T)$ is now a function of $T$. We have assumed that $\langle \hat A \rangle_0 = 0$ throughout
%%%%%%%%%%%%%%%%%%%%%%%%%%%%%%%%%%%%%%%%%%%%%%
%%%%%%%%%%%%%%%%%%%%%%%%%%%%%%%%%%%%%%%%%%%%%%
%%%%%%%%%%%%%%%%%%%%%%%%%%%%%%%%%%%%%%%%%%%%%%
%%%%%%%%%%%%%%%%%%%%%%%%%%%%%%%%%%%%%%%%%%%%%%
\section{Comparison between exact diagonalization and the mean field approximation for computing the superradiant phase boundary of an Ising spin model}
\label{comparisonphasediagram}
In Figure 2 of the main text, we study the different phases that a hybrid spins-cavity system can host when a direct Ising nearest-neighbours interaction between spins $1/2$ in a 1D model is considered. We also assume for simplicity that the coupling of the spins to the cavity is uniform, $\lambda_i \equiv \lambda$. 

\begin{figure}
    \centering
    \includegraphics[width = 0.5\textwidth]{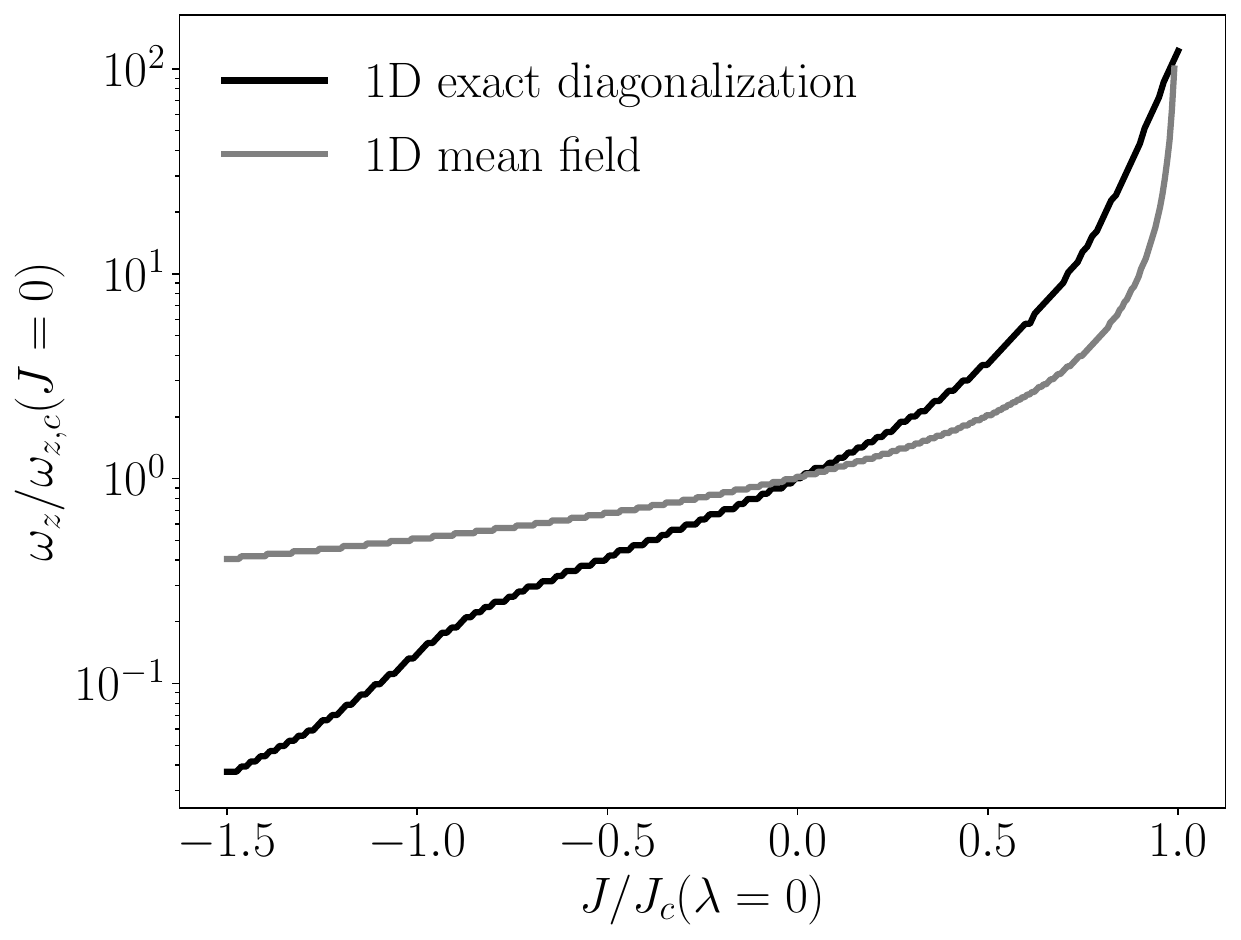}
    \caption{Superradiant phase boundary of the cavity-spin Hamiltonian for $S = 1/2$ spins with ${\cal H}_S = \mathcal H_{\rm Ising}$. Comparison between the boundaries obtained with the mean field method and with exact diagonalization of the 1D Ising Hamiltonian with PBC. The exact diagonalization was performed using a Lanczos routine with 14 spins and a Lanczos basis size of 60. For the units, $J_c (\lambda = 0)$ is the critical Ising coupling in the absence of cavity (which differs by a factor 2 between the mean field approach and exact diagonalization) and $\omega_{z, c} (J = 0)$ is the critical spin frequency in the absence of direct Ising coupling. The parameters used are $\rho = 5.1 \cdot 10^{20} \ {\rm cm}^{-3}$, $\Omega = 1.4 \times 10^9 \ {\rm s}^{-1}$ and $\nu = 1$.}
    \label{fig:ising_phasediagram_comparison}
\end{figure}

Two approaches are then possible to compute the phase boundary between the superradiant and normal phases. From a practical viewpoint, it is convenient to perform the mean field approximation on the direct Ising interaction part of the Hamiltonian. Since the cavity-mediated Ising interaction is all-to-all, it can be exactly treated with mean field.  Proceeding in this fashion from Eq. (4) in the main text (with $\hbar = 1)$, we obtain 
\begin{equation}
    \mathcal{H}_{\mathrm{eff}}^{\mathrm{MF}}=\frac{\omega_{z}}{2} \sum_{j} \sigma_{j}^{z}-\frac{2 \lambda^{\prime 2}}{\Omega} m_{x} \sum_{j} \sigma_{j}^{x}+\frac{\lambda^{\prime 2} N}{\Omega} m_{x}^{2} \,.
    \label{eq:HeffMFIsing}
\end{equation}
Which is the mean field effective Hamiltonian of a Dicke model with a modified light-matter coupling
\begin{equation}
    \lambda^{\prime 2}=\lambda^{2}+\frac{J \omega_{c}}{2} \,.
\end{equation}
From Eq. \ref{eq:HeffMFIsing} one obtains an analytical critical condition via free energy minimization with respect to the variational parameter $m_x$
\begin{equation}
    \lambda'^2=\frac{1}{4} \Omega \omega_{z} \operatorname{coth}\left(\beta \frac{\omega_{z}}{2}\right) \,.
\end{equation}
An alternative calculation simply resorts to the critical condition given by Eq. (5) in the main text, which requires the computation of the response function $\mathcal R$. In the case of homogeneous coupling, the latter is proportional to the magnetic susceptibility in the $xy$-plane, $\chi_\perp$. This can be computed numerically by introducing a perturbative field along the $x$-direction in the Ising Hamiltonian
\begin{equation}
    \mathcal H_S = \mathcal{H}_{\rm Ising} + \epsilon \sum_j \sigma_j^x = \frac{\omega_z}{2} \sum_j \sigma_j^z - \frac{J}{2} \sum_i \sigma_i^x \sigma_{i+1}^x + \epsilon \sum_j \sigma_j^x  \,.
\end{equation}
The susceptibility is then defined as $\expval{\sigma^x} = \expval{\sigma^x}_0 + \epsilon \chi_\perp$. With a Lanczos routine, we perform exact diagonalization of the Hamiltonian to compute $\expval{\sigma^x}$ and $\expval{\sigma^x}_0$. In Fig. \ref{fig:ising_phasediagram_comparison} we compare the phase boundaries obtained with the two methods here described.  We find that the mean field approximation provides an excelent qualitative description of the sinergy between intrinsic and cavity-mediated Ising interactions. 

\section{Expressing the effective spin-spin interaction in terms of the root mean square coupling $\bar \lambda$}
\label{rootmeansquare}
Consider the cavity-mediated effective coupling that appears in the effective spin Hamiltonian $\mathcal H_S$,
\begin{equation}
    \mathcal H_{\rm eff,int} = -\frac{1}{\hbar \Omega} \left[\sum_j \frac{\hbar \lambda_j}{\sqrt{N}}\left( e^{i \theta_j} S^+_j  + {\rm h.c.} \right) \right]^2 \, .
\end{equation}
Without loss of generality, let us focus on the case that $\theta_j = 0$ and apply a mean field approximation to the spins $S^x_j = m_x + \delta(S^x_j)_j$, yielding
\begin{equation}
    \mathcal H_{\rm eff, int}^{\rm MF} = \frac{m_x^2}{\hbar \Omega N} \left(\sum_j \hbar \lambda_j \right)^2 - \frac{2 m_x}{\hbar \Omega N} \sum_j \hbar \lambda_j S_j^x \left(\sum_i \hbar \lambda_i \right) \, .
\end{equation}
The Cauchy–Schwarz inequality teaches us that the root mean square can be used as an upper bound for the arithmetic mean $\bar \lambda^2 = N^{-1} \sum_j \lambda_j^2 \geq N^{-2} \left(\sum_j \lambda_j\right)^2$. Considering that the ${\lambda_j}$ are definite positive, this upper bound is reasonably close to the actual value, allowing us to approximate $\mathcal H_{\rm eff, int}^{\rm MF}$ by
\begin{equation}
    \mathcal H_{\rm eff, int}^{\rm MF} = \frac{\hbar m_x^2 \bar  \lambda^2 N}{\Omega}  - \frac{\hbar 2 m_x \bar \lambda}{\Omega} \sum_j \lambda_j S_j^x \, .
\end{equation}
Benefiting again from the bound, let us perform a mean field approximation on the ${\lambda_j}$ where we neglect fluctuations around the root mean square $\lambda_j = \bar \lambda + \delta(\lambda_j) \simeq \bar \lambda $. With this approximation, we can finally write
\begin{equation}
    \mathcal H_{\rm eff, int}^{\rm MF} = \frac{\hbar m_x^2 \bar  \lambda^2 N}{\Omega}  - \frac{\hbar 2 m_x \bar \lambda^2}{\Omega} \sum_j S_j^x \, .
\end{equation}
Having shown that $\bar \lambda^2$ is a function of only the density of spins $\rho$ and the filling factor $\nu$, this approximation allows us to study the modification of magnetic properties depending exclusively on $\rho$ and $\nu$, as is exemplified in the main text.
\section{Computing the transmission in a driven Dicke model}
\label{transmission}
\subsection{Phenomenological damping - Local master equation}
For simplicity, consider a spin $1/2$ Dicke model coupled to a transmission line ($\hbar = 1$), in which the coupling is modulated by a periodic driving of the cavity
\begin{equation}
    \mathcal H = \frac{\omega_z}{2} \sum_j \sigma_j^z + \Omega a^\dagger a + \frac{\lambda}{\sqrt{N}} (a + a^\dagger) \sum_j \sigma_j^x + f(t) (a + a^\dagger) \, ,
\end{equation}
where $f(t)  = \gamma_1 e^{-i \omega t}$. The quantity to be measured in the experiment is the transmission, which is proportional to the susceptibility of the system to the periodic driving that is the excitation.
The expression for the transmission is $t = -i \gamma_1 \chi_a$. So we are interested in computing $\chi_a$. For that, it is convenient to express the Hamiltonian in terms of total spin operators
\begin{equation}
    \mathcal H = \omega_z S_z + \Omega a^\dagger a + \frac{2 \lambda}{\sqrt{N}} (a + a^\dagger) S_x + f(t) (a + a^\dagger) \,,
\end{equation}
so we can write the equations of motion of the system in terms of these new observables. Presumably, the experiment would be carried out in a regime where the temperature can be approximated by zero, which simplifies the calculation. The equations of motion are the Heisenberg equations of motion ($\dot{\hat O} = i [\mathcal{\hat H}, \hat O]$) plus a phenomenological damping originating from the coupling to the transmission line
\begin{equation}
    \begin{cases}
    \dot S_z = \frac{2 \lambda}{\sqrt{N}} (a + a^\dagger) S_y - \gamma_2 (S_z + N) \,, \\
    \dot S_x = - \omega_z S_y - \gamma_2 S_x \,, \\
    \dot S_y = \omega_z S_x - \frac{2 \lambda}{\sqrt{N}} (a^\dagger + a)S_z - \gamma_2 S_y \,, \\
    \dot a = - i\Omega a - i \frac{2 \lambda}{\sqrt{N}} S_x - i f(t) - \gamma_1 a \,.
    \end{cases}
\end{equation}
Here $\gamma_1$ and $\gamma_2$ are, respectively, the damping constants for the cavity and the spins. These can be formally obtained by considering each subsystem (spins and cavity) in isolation, and then postulating for each a more general model where the subsystem is coupled to a bosonic environment that acts as a heat reservoir. The dynamics of each each subsystem are then given by their so-called local master equations, in which the bath is traced-out, leaving the dissipation terms that we introduce here. The denomination \textit{local} stems from the fact that the dissipation terms are computed for each subsystem independently but then introduced into the equations of motion of the joint system. Below we contrast this approach with the more consistent global master equation, in which a single master equation is written for the joint system. The dissipation terms are computed taking into account the complete system, instead of the isolated subsystems. In many instances, the phenomenological damping given by the local approach provides an accurate description of the dissipative dynamics. However, in cases where the intersubsystem coupling is strong, the significant modification to the energy levels warrants the use of the global master equation \cite{Hofer_2017}. At the very least, to validate the results obtained with the simpler local master equation. If we consider $f(t)$ as a perturbation,  i.e. $\gamma_1 \ll \Omega$, we can write the average values of the operators in terms of their unperturbed values and the corresponding susceptibility: $\expval{S_\mu} = \expval{S_\mu}_0 + \gamma_1 e^{-i \omega t} \chi_\mu$, $\expval{a} = \alpha_0 + \gamma_1 e^{-i \omega t} \chi_a$. Imposing the equilibrium condition: $\dot S_\mu = \dot a = 0$, and assume $\expval{a} = \expval{a
^\dagger}$ to be real, we arrive to the system of equations that must be solved in order to obtain $\chi_a$
\begin{equation}
    \begin{cases}
    -i \omega \chi_z = \frac{4 \lambda}{\sqrt{N}} \left(\chi_y \alpha_0 + \expval{S_y}_0 \chi_a \right) - \gamma_2 \chi_z \,, \\
    -i \omega \chi_x = - \omega_z \chi_y - \gamma_2 \chi_x \,, \\
    -i \omega \chi_y = \omega_z \chi_x - \frac{4\lambda}{\sqrt{N}} \left(\chi_z \alpha_0 + \expval{S_z}_0 \chi_a \right) - \gamma_2 \chi_y \,, \\
    -i \omega \chi_a = - i \Omega \chi_a - i \frac{2 \lambda}{\sqrt{N}} \chi_x - i - \gamma_1 \chi_a \,.
    \end{cases}
\end{equation}
Solving for $\chi_a$ yields
\begin{equation}
    \chi_a = \frac{1}{(\omega - \Omega) + i \gamma_1 + \frac{8 \lambda^2 \omega_z \expval{S_z}_0}{N} \left((\omega + i \gamma_2)^2 - \omega_z^2 - \frac{16 \lambda^2 \alpha_0^2}{N} \right)^{-1}} \, .
\end{equation}
Noting that $\expval{S_\mu}_0 = N \expval{\sigma_\mu}_0 / 2$ allows us to write
\begin{equation}
\label{eq:tw}
    t = - \frac{i \gamma_1}{(\omega - \Omega) + i \gamma_1 + 4 \lambda^2 \chi_{\perp} (\omega)} \, ,
\end{equation}
with
\begin{equation}
    \chi_\perp(\omega) = \frac{\omega_z \expval{\sigma_z}_0}{(\omega - i \gamma_2)^2 - \omega_z^2 - \frac{16 \lambda^4 \expval{\sigma_x}_0^2}{\Omega^2}} \,.
\end{equation}
The dependece on the normal and superradiant phases is through the equilibrium values of $\expval{\sigma_z}_0$ and $\expval{\sigma_x}_0$.

\subsection{Global master equation}
Introducing the total spin operators $S^z = \sum_j \frac{\sigma^z_j}{2}$, the Hamiltonian of the spins-cavity system is
\begin{equation}
    \mathcal H = \omega_z S^z + \Omega a^\dagger a + \frac{2 \lambda}{\sqrt{N}} (a + a^\dagger) S^x \,.
\end{equation}
We use a Holstein-Primakoff transformation on the spins to switch to a bosonic representation, where the system consists of two coupled oscillators. When applying the transformation, we must distinguish two cases, when the system is in the normal phase and when it is superradiant. In the normal phase at zero temperature, where the ground state is an empty cavity with relaxed spins, we use the relations
\begin{equation}
    S^z = -\frac{N}{2} + s^\dagger s \, ; \qquad S^+ = \sqrt{N} s^\dagger \, .
\end{equation}
In the thermodynamic limit, $N \to \infty$, these relations are exact. The new operators $s^\dagger$, $s$ obey the bosonic commutation relations and hence constitute bosonic creation and annihilation operators, respectively. The resulting Hamiltonian
\begin{equation}
    \mathcal H = \omega_z s^\dagger s + \Omega a^\dagger a + \lambda (s + s^\dagger)(a + a^\dagger) \, ,
\end{equation}
describes two coupled harmonic oscillators. In the theory of open quantum systems, dissipation is introduced by expanding the system to include bosonic baths that couple to each of the system's degrees of freedom. In this case, to each of the oscillators. Henceforth, we follow the prescription of Breuer and Petruccione \cite{BRE02}.  We assume the coupling operators to be $a + a^\dagger$ and $s + s^\dagger$ for each subsystem. The master equation formalism requires that we find the spectral decomposition of the coupling operators in terms of the eigenoperators of the Hamiltonian. We diagonalize the Hamiltonian with the use of the Bogoliubov transformation.
It can be written in matrix form as $\mathcal H = \frac{1}{2} \phi^\dagger M \phi $, with $\phi^t = (a, s, a^\dagger, s^\dagger)$ and 
\begin{equation}
M = 
    \begin{pmatrix}
\Omega & \lambda & 0 & \lambda \\
\lambda & \omega_z & \lambda & 0 \\
0 & \lambda & \Omega & \lambda \\
\lambda & 0 & \lambda & \omega_z 
\end{pmatrix} \,.
\end{equation}
A Bogoliubov transformation defines a new set of bosonic operators $\psi = T^{-1} \phi$ such that the Hamiltonian can be cast in diagonal form $\mathcal H = \frac{1}{2} \psi^\dagger M' \psi$, with $M' = T^\dagger M T$ a diagonal matrix. Imposing canonical commutation relations on the transformed operators, i.e. that the transformed vector has the structure $\phi^t = (d_1, d_2, d_1^\dagger, d_2^\dagger)$, yields a condition of pseudounitarity for the transformation matrix $T^\dagger S_3 T = S_3$, where $S_3 = \rm{diag}(1, 1, -1, -1)$. Furthermore, it implies that $S_3 M' = T^{-1} S_3 M T$. So, in practice, performing the transformation amounts to finding the eigenvalues and eigenvectors of $S_3 M$ and making sure that the eigenvectors are pseudonormalized so as to satisfy the condition of pseudounitarity of the change-of-basis matrix $T$. The resulting Hamiltonian can be written as 
\begin{equation}
    \mathcal H = \omega_1 d_1^\dagger d_1 + \omega_2 d_2^\dagger d_2,
\end{equation}
where $\omega_1$, $\omega_2$ are the positive eigenvalues of $S_3 M$. We assign $\omega_1$ to the mode that goes to zero at the transition: $\omega_1(\lambda_c) = 0$. This softening of one of the modes is the critical behaviour of the system. The relation between the old and new bosonic operators can be written as
\begin{align}
    &a = g_1 d_1 + g_2 d_2 + g_3 d_1^\dagger + g_4 d_2^\dagger \, , \label{eq:bogoliubov_a}\\
    &s = h_1 d_1 + h_2 d_2 + h_3 d_1^\dagger + h_4 d_2^\dagger \, .
\end{align}
Defining the coefficients $G_1 = g_1 + g_3^*$, $G_2 = g_2 + g_4^*$,
$H_1 = h_1 + h_3^*$ and $H_2 = h_2 + h_4^*$, we can write the coupling operators in terms of the new bosonic operators
\begin{align}
    &a + a^\dagger = G_1 d_1 + G_2 d_2 + h.c. \, \\ 
    &s + s^\dagger = H_1 d_1 + H_2 d_2 + h.c. \,.
\end{align}
After performing the customary Born-Markov and secular approximations, we arrive at the master equation for the density matrix $\rho_S$ of the system of two coupled oscillators
\begin{equation}
    \begin{split}
        \frac{d}{dt} \rho_S = -i [\mathcal H, \rho_S] &+ \gamma_1(\omega_1) |G_1|^2 \left[d_1 \rho_S d_1^\dagger - \frac{1}{2} \left\{d_1^\dagger d_1, \rho_S \right\} \right] \\
        &+\gamma_1(\omega_2) |G_2|^2 \left[d_2 \rho_S d_2^\dagger - \frac{1}{2} \left\{d_2^\dagger d_2, \rho_S \right\} \right] \\
        &+\gamma_1(-\omega_1) |G_1|^2 \left[d_1^\dagger \rho_S d_1 - \frac{1}{2} \left\{d_1 d_1^\dagger, \rho_S \right\} \right] \\
        &+\gamma_1(-\omega_2) |G_2|^2 \left[d_2^\dagger \rho_S d_2 - \frac{1}{2} \left\{d_2 d_2^\dagger, \rho_S \right\} \right] \\
        &+\gamma_2(\omega_1) |H_1|^2 \left[d_1 \rho_S d_1^\dagger - \frac{1}{2} \left\{d_1^\dagger d_1, \rho_S \right\} \right] \\
        &+\gamma_2(\omega_2) |H_2|^2 \left[d_2 \rho_S d_2^\dagger - \frac{1}{2} \left\{d_2^\dagger d_2, \rho_S \right\} \right] \\
        &+\gamma_2(-\omega_1) |H_1|^2 \left[d_1^\dagger \rho_S d_1 - \frac{1}{2} \left\{d_1 d_1^\dagger, \rho_S \right\} \right] \\
        &+\gamma_2(-\omega_2) |H_2|^2 \left[d_2^\dagger \rho_S d_2 - \frac{1}{2} \left\{d_2 d_2^\dagger, \rho_S \right\} \right] \, .
    \end{split}
\end{equation}
Note that we have neglected the Lamb shift. It must be noted that the secular approximation breaks down at the transition, where $\omega_1 \to 0$. Four (two for each bath) of the rotating terms that are neglected by the secular equation become stationary at this point, thus surviving the approximation. These are
\begin{equation}
    \begin{split}
        &\gamma_1(0) G_1^2 \left[d_1 \rho_S d_1 -  \frac{1}{2} \left\{d_1 d_1, \rho_S \right\}\right] \\
        + \; &\gamma_1(0) (G_1^*)^2 \left[d_1^\dagger \rho_S d_1^\dagger -  \frac{1}{2} \left\{d_1^\dagger d_1^\dagger, \rho_S \right\}\right] \\
        + \, &\gamma_2(0) H_1^2 \left[d_1 \rho_S d_1 -  \frac{1}{2} \left\{d_1 d_1, \rho_S \right\}\right] \\
        + \; &\gamma_2(0) (H_1^*)^2 \left[d_1^\dagger \rho_S d_1^\dagger -  \frac{1}{2} \left\{d_1^\dagger d_1^\dagger, \rho_S \right\}\right] \,.
    \end{split}
\end{equation}%
To compute the susceptibility $\chi_\alpha$, we recall the Bogoliubov decomposition of operator $a$ \eqref{eq:bogoliubov_a}, from which it is clear that $\chi_\alpha = g_1 \chi_1 + g_2 \chi_2 + g_3 \chi_1^\dagger + g_4 \chi_2^\dagger$. Note that the susceptibilities are scalars and thus the $\dagger$ sign is not used indicate hermitian conjugacy, but simply to mark that $\chi_1^\dagger$ ($\chi_2^\dagger$) is the susceptibility of operator $d_1^\dagger$ ($d_2^\dagger$). Note as well that $\chi_1^\dagger \neq \chi_1^*$. We are thus interested in the dynamics of the transformed operators. Focusing on $d_1$, we find that the dynamics are completely uncoupled from $d_2$, with master equation
\begin{equation}
   \begin{split}
        \dot d_1 = i \left[\mathcal H, d_1 \right] &+ \left(\gamma_1(\omega_1) |G_1|^2 + \gamma_2(\omega_1) |H_1|^2 \right) \left[d_1^\dagger d_1 d_1 - \frac{1}{2} \left\{d_1^\dagger d_1, d_1 \right\} \right] \label{eq:first_master_d1} \\
        &+ \left(\gamma_1(-\omega_1) |G_1|^2 + \gamma_2(-\omega_1) |H_1|^2 \right) \left[d_1 d_1 d_1^\dagger - \frac{1}{2} \left\{d_1 d_1^\dagger, d_1 \right\} \right] \, .
   \end{split}
\end{equation}
As for the terms that survive the secular approximation at the transition, we have
\begin{equation}
    \begin{split}
        &\left(\gamma_1(0) G_1^2 + \gamma_2(0) H_1^2 \right) \left[d_1^\dagger d_1 d_1^\dagger - \frac{1}{2} \left\{d_1^\dagger d_1^\dagger, d_1 \right\} \right] \\
        + \; & \left(\gamma_1(0) (G_1^*)^2 + \gamma_2(0) (H_1^*)^2 \right) \left[d_1 d_1 d_1 - \frac{1}{2} \left\{d_1 d_1, d_1 \right\} \right] = 0 \, .
    \end{split}
\end{equation}
They do not contribute to the dynamics. Assuming a flat bath $\gamma_j(\omega) = 2 \gamma_j N(\omega)$, with $N(\omega)$ the bosonic ocupation number at temperature $T$ of the bosonic bath, we obtain a simplification for Eq. \eqref{eq:first_master_d1} valid at all temperatures, and particularly at $T = 0$,
\begin{equation}
    \dot d_1 = i \left[\mathcal H, d_1 \right] + \left(\gamma_1 |G_1|^2 + \gamma_2 |H_1|^2 \right) d_1 \,.
    \label{eq:master_d1}
\end{equation}
Analogusly, we obtain for $d_2$
\begin{equation}
    \dot d_2 = i \left[\mathcal H, d_2 \right] + \left(\gamma_1 |G_2|^2 + \gamma_2 |H_2|^2 \right) d_2 \,.
\end{equation}
In order to compute $\chi_\alpha$, we introduced a periodic perturbation in $\mathcal H$ of the form $\gamma_1 e^{-i \omega t} (a + a^\dagger)$ in the local master equation formalism. Having switched to new bosonic operators, this translates to
\begin{equation}
    \mathcal H = \omega_1 d_1^\dagger d_1 +  \omega_2 d_2^\dagger d_2 + \epsilon \gamma_1 e^{-i \omega t} \left(|G_1|^2 d_1 + |G_2|^2 d_2 + |G_1|^2 d_1^\dagger + |G_2|^2 d_2^\dagger   \right) \,.
    \label{eq:perturbed_ham}
\end{equation}
We define the susceptibilities as the response of each operator to this perturbation, for instance $\expval{d_1} = \expval{d_1}_0 + \gamma_1 e^{-i \omega t} \chi_1$, with $\expval{d_1}_0$ the equilibrium value. Introducing this and the perturbed Hamiltonian \eqref{eq:perturbed_ham} in Eq. \ref{eq:master_d1} we obtain
\begin{equation}
    \chi_1 = \frac{ |G_1|^2}{(\omega - \omega_1) + i \left(\gamma_1 |G_1|^2 + \gamma_2 |H_1|^2 \right)} \,.
\end{equation}
Proceeding analogously, we obtain the remaining susceptibilities
\begin{align}
    &\chi_2 = \frac{|G_2|^2}{(\omega - \omega_2) + i \left(\gamma_1|G_2|^2 + \gamma_2 |H_2|^2 \right)} \,, \\
    &\chi_1^\dagger = -\frac{ |G_1|^2}{(\omega + \omega_1) + i  \left(\gamma_1 |G_1|^2 + \gamma_2 |H_1|^2 \right)} \,, \\
     &\chi_2^\dagger = -  \frac{|G_2|^2}{(\omega + \omega_2) + i  \left(\gamma_1 |G_2|^2 + \gamma_2 |H_2|^2 \right)} \,.
\end{align}
Finally, the transmission is $t = - i \gamma_1 \chi_\alpha$, or, written in terms of the new susceptibilities,
\begin{equation}
    t = -i \gamma_1 \left(g_1 \chi_1 + g_2 \chi_2 + g_3 \chi_1^\dagger + g_4 \chi_2^\dagger \right) \,.
    \label{eq:transmission}
\end{equation}
It is convenient to recall the range of validity of the expression for the transmission just presented. The use of the Holstein-Primakoff transformation implies that we stay on the normal phase at zero temperature, where the equilibrium state consists entirely of relaxed spins. We can, however, extend it to finite temperatures with the use of a clever observation. It can be seen from the critical condition $4 \lambda^2 = \Omega \omega_z \coth \left(\beta \omega_z / 2 \right)$ that we can define an effective temperature-dependent coupling $\lambda^2 (T) = \lambda^2 \tanh \left(\beta \omega_z / 2 \right)$. Realizing that $|\expval{\sigma_z}_0| = \tanh \left(\beta \omega_z / 2 \right)$, we see that as the temperature is increased, $|\expval{\sigma_z}_0|$ decreases: a growing fraction of the spins begin to fluctuate thermally, leaving a lower number in their ground state, where they are free to respond to the magnetic field generated by the cavity. In the model under study, the coupling is also related to the density of spins as $\lambda^2 \propto \rho$. Essentially, this means that we can mimic the effect of temperature on the system by working with a system at zero temperature where the density of spins is decreased by a factor $\tanh \left(\beta \omega_z / 2 \right)$. This allows us to recycle Eq. \eqref{eq:transmission} for the finite temperature regime.

Let us return now to the Holstein-Primakoff transformation. The result obtained thus far is only valid in the normal phase, where the ground state is the trivial one. The superradiant phase requires more care. The new bosonic operators must reproduce the superradiant ground state, which contains excited spins as well as a photonic population. In practice, this is achieved with a regular Holstein-Primakoff transformation plus a displacement of the bosonic operators \cite{clive2003}. The resulting Hamiltonian, again exact in the thermodynamic limit, reads
\begin{equation}
\bar{\mathcal  H}= \Omega \bar a^{\dagger} \bar a + \frac{\omega_z}{2 \mu}(1+\mu) \bar s^{\dagger} \bar s + \frac{\omega_z(1-\mu)(3+\mu)}{8 \mu(1+\mu)}\left(\bar s^{\dagger} + \bar s\right)^{2} + \lambda \mu \sqrt{\frac{2}{1+\mu}}\left(\bar a^{\dagger} + \bar a\right)\left(\bar s^{\dagger} + \bar s\right) \,,
\end{equation}
Where $\mu = \omega_z \Omega / (4 \lambda^2)$ and $\bar a$, $\bar s$ are the displaced bosonic operators. We omit the expressions for the displacements as they are irrelevant in the derivation of the transmission. The driving field suffers a trivial displacement that does not affect the susceptibility. The idea behind the displacement is to couple the dissipative baths to the displaced bosons, such that when these are asymptotically carried to zero occupation by the dissipation, the original bosons retain their correct equilibrium occupation of the superradiant phase. So with this newly defined Hamiltonian, one proceeds analogously to the normal-phase case, the only change is in the spectral decomposition of the new coupling operators $\bar a + \bar a^\dagger$ and $\bar s + \bar s^\dagger$, which originates from the fact that we must now diagonalize 
\begin{equation}
\bar M = 
    \begin{pmatrix}
\Omega & g & 0 & \lambda \\
g & \tilde \omega_z + \epsilon & g & \epsilon \\
0 & g & \Omega & g \\
g & \epsilon & g & \tilde \omega_z + \epsilon
\end{pmatrix} \,.
\end{equation}
Here $g = \lambda \mu \sqrt{\frac{2}{1+\mu}}$, $\tilde \omega_z = \frac{\omega_z}{2 \mu}(1+\mu)$ and $\epsilon = \frac{\omega_z(1-\mu)(3+\mu)}{8 \mu(1+\mu)}$. So Eq. \eqref{eq:transmission} is still valid in the superradiant regime, but with the caveat that now $\omega_1$, $\omega_2$, $\{g_i\}$ and $\{h_i\}$ are given by the Bogoliubov transformation of the new Hamiltonian given by $\bar M$.

\subsection{Comparing the two approaches}

The Bogoliubov transformation used in the derivation of the global master equation must, in practice, be computed numerically. From there on, it is just a matter of using the obtained values of the coefficients $\{g_i\}$, $\{h_i\}$ and the eigenenergies $\omega_1$, $\omega_2$ to compute the susceptibilities and the transmission. Nevertheless, it is still possible to study the limit of low $\omega$ and $\omega_z \sim \omega_{\rm z, c} (T = 0)$ analytically.

\begin{figure}
    \centering
    \includegraphics[width = 0.5\textwidth]{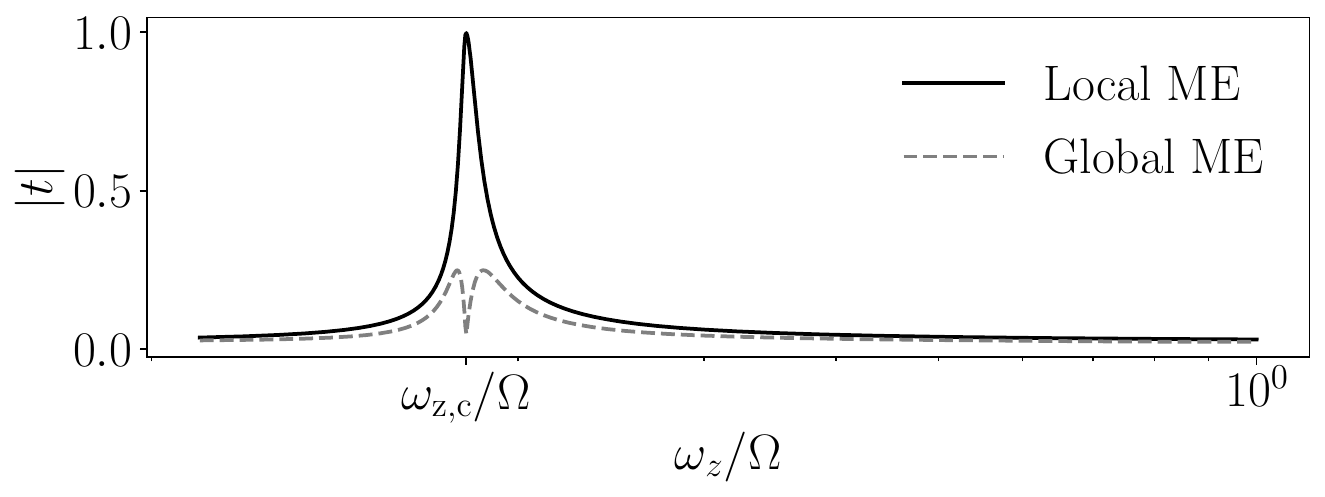}
    \caption{Transmission profiles of the spin 1/2 Dicke model for $\omega \to 0$ at zero temperature. We compare the results obatined with the local master equation (solid black) and global master equation (dashed grey). The parameters used are  $\gamma_1 = 0.025 \, \Omega$, $\gamma_2 = 0.025 \, \omega_z$, $ \rho = 5.1 \times 10^{20} \, {\rm cm}^{-3}$, $\Omega = 1.4 \times 10^9 \ {\rm s}^{-1}$ and $\nu = 0.25$. }
    \label{fig:transmission_extra}
\end{figure}

Within the local approach, at $\omega = 0$ we have
\begin{equation}
    \chi_\perp = \frac{\omega_z^{-1}}{\gamma_2^2 + \omega_z^2} \approx \frac{1}{\omega_z} \,.
\end{equation}
At the critical point, $4 \lambda^2 = \omega_z \Omega$, this yields
\begin{equation}
    t = -\frac{i \gamma_1}{-\Omega + i \gamma_1 + \Omega} = -1 \,.
\end{equation}
We can contrast this result with the one obtained with the global master equation. At the critical point, we find that $\omega_1 = 0$, so for $\omega = 0$, both $\chi_1$ and $\chi_1^\dagger$ will be resonant. With values
\begin{align}
    &\chi_1 = \frac{|G_1|^2}{i \left(\gamma_1 |G_1|^2 + \gamma_2 |H_1|^2 \right)} \,, \\
    &\chi_1^\dagger = -\frac{|G_1|^2}{i \left(\gamma_1 |G_1|^2 + \gamma_2 |H_1|^2 \right)} \,.
\end{align}
So the resulting transmission is given by
\begin{equation}
    t = -\frac{i \gamma_1 |G_1|^2}{i\left(\gamma_1 |G_1|^2 + \gamma_2 |H_1|^2 \right)} \left(g_1 - g_3 \right)
\end{equation}
In this case the result is not as clear as with the local approach as it depends on the coefficients of the Bogoliubov transformation. To gain further insight, in Figure \ref{fig:transmission_extra} we plot the transmission in the low $\omega$ regime as a function of $\omega_{\rm z}$. Note that this is at $T = 0$, where a phase transition is expected.
What we find is that $(g_1- g_3)$, and subsequently the global transmission, goes to zero right at the transition, despite resonant behaviour right before and after the critical point. This contrasts with the result obtained for the local transmission, but in any case, both approaches show a signature either at (local) or immediately around (global) the phase transition.

%========== Bibliography =============
\bibliography{sm.bib}